\documentclass[10pt,twocolumn,showpacs,amsmath,amssymb,aps,prx,longbibliography,superscriptaddress]{revtex4-1}
\usepackage{times}
\usepackage[utf8]{inputenc}
\usepackage{color}
\usepackage{array}
\usepackage{verbatim}
\usepackage{multirow}
\usepackage{amsmath}
\usepackage{amssymb}
\usepackage{graphicx}
\usepackage{esint}
\usepackage[unicode=true,
 bookmarks=true,bookmarksnumbered=true,bookmarksopen=false,
 breaklinks=true,pdfborder={0 0 0},backref=false,colorlinks=true]
 {hyperref}
\usepackage{enumitem}
\usepackage{cleveref}
\usepackage{xcolor}
\usepackage[bb=boondox]{mathalfa}
\usepackage[title]{appendix}
\usepackage{makecell}

\hypersetup{
    linkcolor=red,          % color of internal links (change box color with linkbordercolor)
    citecolor=blue,        % color of links to bibliography
    filecolor=magenta,      % color of file links
    urlcolor=magenta           % color of external links
}

\usepackage{cleveref}
\crefname{appendix}{Appendix}{Appendices}
\crefname{equation}{Eq.}{Eqs.}
\crefname{figure}{Fig.}{Figs.}
\crefname{table}{Table}{Tables}
\crefname{section}{Sec.}{Secs.}

\renewcommand{\paragraph}[1]{\vspace{0.2cm}{\bf \textit{#1}}}
\def\ie{{\it i.e.},\ }
\def\eg{{\it e.g.},\ }
\def\etc{{\it etc.}}

\definecolor{Gray}{gray}{0.85}
\newcolumntype{a}{>{\columncolor{Gray}}c}

\usepackage{simpler-wick}
\allowdisplaybreaks[1] % page breaks are allowed, but avoided if possible

\newcommand{\mrm}{\mathrm}

\newcommand{\td}{\widetilde}

\newcommand{\hH}{\hat{H}}

\def\pare#1{\left( #1 \right)}
\def\brak#1{\left[#1\right]}

\def\bra#1{\langle #1 |}
\def\ket#1{| #1 \rangle}

\def\inn#1{\langle #1 \rangle}

\def\abs#1{\left| #1 \right|}
\def\Im{\mathrm{Im}}

\def\tr{\mathrm{Tr}}
\def\up{\uparrow}
\def\down{\downarrow}

\def\kk{ {\mathbf{k}} }
\def\RR{\mathbf{R}} 
\def\qq{\mathbf{q}}
\def\QQ{\mathbf{Q}}
\def\oo{\mathbf{0}}
\def\GG{\mathbf{G}}

\def\ee{\epsilon}
\def\NN{\mathcal{N}}
\def\eco{\epsilon_{c,1}}
\def\ect{\epsilon_{c,3}}
\def\eef{\epsilon_f}
\def\tk{T_{\rm K}}

\begin{document}
\title{Kondo Phase in Twisted Bilayer Graphene}

\author{Geng-Dong Zhou}
\affiliation{International Center for Quantum Materials, School of Physics, Peking University, Beijing 100871, China}

\author{Yi-Jie Wang}
\affiliation{International Center for Quantum Materials, School of Physics, Peking University, Beijing 100871, China}

\author{Ninghua Tong}
\affiliation{Department of Physics, Renmin University of China, Beijing 100872, China}

\author{Zhi-Da Song}
\email{songzd@pku.edu.cn}
\affiliation{International Center for Quantum Materials, School of Physics, Peking University, Beijing 100871, China}

\date{\today}

\begin{abstract}
A number of interesting physical phenomena have been discovered in magic-angle twisted bilayer graphene (MATBG), such as superconductivity, correlated gapped and gapless phases, {\it etc}. The gapped phases are believed to be symmetry-breaking states described by mean-field theories, whereas gapless phases exhibit features not explained by mean-field theories. 
This work, using a combination of poor man's scaling, numerical renormalization group, and dynamic mean-field theory, demonstrates that the gapless phases are the heavy Fermi liquid state, where some symmetries might be broken while the others are preserved. We adopt the recently proposed topological heavy fermion model for MATBG, where effective local orbitals around AA-stacking regions and Dirac fermions surrounding them play the roles of local moments (LM's) and itinerant electrons, respectively. At zero temperature and most non-integer fillings, the ground states are found to be heavy Fermi liquids and exhibit Kondo resonance peaks. 
The Kondo temperature $\tk$ is found at the order of 1meV. 
A higher temperature than $\tk$ will drive the system into a metallic LM phase where disordered LM's, obeying Curie's law, and a Fermi liquid formed by itinerant electrons coexist.
At integer fillings $\pm1,\pm2$, $\tk$ is suppressed to zero or a value weaker than the RKKY interaction, leading to Mott insulators or symmetry-breaking states. 
Remarkably, this theory offers a unified explanation for several experimental observations, such as zero-energy peaks and quantum-dot-like behaviors in STM, the so-called Pomeranchuk effect, and the saw-tooth feature of inverse compressibility, {\it etc.}
For future experimental verification, we predict that the Fermi surface in the gapless phase will shrink upon heating - as a characteristic of the heavy Fermi liquid. We also conjecture that the heavy Fermi liquid is the parent state of the observed unconventional superconductivity because the Kondo screening reduces the overwhelming Coulomb interaction ($U\sim 60$meV) to a rather small residual effective interaction ($U^*\sim 1$meV) that is comparable to possible weak attractive interactions.  
\end{abstract}

\maketitle

\section{Introduction}

After the discovery of superconductivity \cite{cao_unconventional_2018} and correlated insulators \cite{cao_correlated_2018} in magic-angle twisted bilayer graphene (MATBG) \cite{bistritzer_moire_2011}, MATBG has become a platform for studying new correlation effects in flat-band systems and has received extensive attention.
Remarkably rich physics, including the interplay between superconductivity \cite{lu_superconductors_2019,yankowitz_tuning_2019,saito_independent_2020,saito_independent_2020,stepanov_untying_2020,liu_tuning_2021,arora_superconductivity_2020,cao_nematicity_2021,oh_evidence_2021} and strong correlation \cite{lu_superconductors_2019,saito_independent_2020,stepanov_untying_2020,liu_tuning_2021,xie_spectroscopic_2019,sharpe_emergent_2019,choi_electronic_2019,kerelsky_maximized_2019,jiang_charge_2019,wong_cascade_2020,zondiner_cascade_2020,choi_interaction-driven_2021}, 
interaction driven Chern insulators \cite{serlin_intrinsic_2020,nuckolls_strongly_2020,choi_correlation-driven_2021,saito_hofstadter_2021,das_symmetry-broken_2021,wu_chern_2021,park_flavour_2021}, 
strange metal behaviors \cite{polshyn_large_2019,cao_strange_2020,jaoui_quantum_2022}, and the Pomeranchuk effect \cite{rozen_entropic_2021,saito_isospin_2021}, \etc, have been observed in MATBG.
Several theoretical understandings of the correlated gapped states have also been achieved: 
The strong correlation arises from the two topological flat bands \cite{bistritzer_moire_2011,song_all_2019,po_faithful_2019,ahn_failure_2019,tarnopolsky_origin_2019,liu_pseudo-landau-level_2019, TBG2}, and each is four-fold degenerate due to the spin and valley d.o.f. 
A large U(4) symmetry group \cite{kang_strong_2019,bultinck_ground_2020,seo_ferromagnetic_2019, TBG3,vafek_renormalization_2020} emerges in the flat-band limit, where the bandwidth is counted as negligible. 
Then the observed correlated gapped states can be understood as flavor polarized states \cite{kang_strong_2019,bultinck_ground_2020,seo_ferromagnetic_2019,vafek_renormalization_2020,TBG4,TBG6,xie_nature_2020,liu_theories_2021,cea_band_2020,venderbos_correlations_2018,ochi_possible_2018,zhang_correlated_2020,liu_nematic_2021,da_liao_correlation-induced_2021,kang_non-abelian_2020,soejima_efficient_2020,hejazi_hybrid_2021,xu_valence_2019,kennes_strong_2018,classen_competing_2019} that spontaneously break the U(4) symmetry.
The continuous U(4) degeneracy also leads to Goldstone mode fluctuations \cite{TBG5,khalaf2020soft}.

Less theoretical understandings have been achieved for the gapless states.
They exhibit some exotic phenomena beyond naive mean-field descriptions: 
(i) zero-energy peaks in spectral density at low temperatures~\cite{nuckolls_strongly_2020,choi_correlation-driven_2021,choi_interaction-driven_2021,oh_evidence_2021}, (ii) a cascade of transitions as that of a quantum dot at higher temperatures~\cite{wong_cascade_2020,choi_interaction-driven_2021}, (iii) the so-called Pomeranchuk effect where local moment (LM) develops upon heating \cite{rozen_entropic_2021,saito_isospin_2021}, (iv) the saw-tooth feature of inverse compressibility~\cite{rozen_entropic_2021,wong_cascade_2020,zondiner_cascade_2020,park_flavour_2021,saito_isospin_2021}. 
These phenomena are not connected or explained on a microscopic level in prior theories.

In this work, we perform systematic analytical and numerical investigations to the recently developed topological heavy fermion (THF) model \cite{song_magic-angle_2022,shi2022heavy} for MATBG. 
The THF model consists of localized $f$-electrons at the AA-stacking regions of MATBG, which form a triangular lattice of LM's, and plane-wave-like itinerant Dirac $c$-electrons, which tend to screen the LM's due to the Kondo effect \cite{nozieres_kondo_1980, tsvelick_exact_1983, coleman_new_1984, bickers_review_1987, affleck_critical_1991, affleck_kondo_1991, emery_mapping_1992, tsvelik_phenomenological_1993, furusaki_kondo_1994, cox_exotic_1998, parcollet_overscreened_1998, coleman2005quantum, kotliar_electronic_2006, werner_continuous-time_2006, gegenwart_quantum_2008, si_heavy_2010, dzero_topological_2010, lu_correlated_2013}. 
We derive a phase diagram consisting of symmetry-breaking states at zero temperature, heavy Fermi liquids at zero temperature, and metallic LM states at finite temperature where disordered LM's and itinerant electrons coexist.   
This phase diagram provide natural explanations for the experiments mentioned in the last paragraph.

For this work to be self-contained, in \cref{sec:THF} we review the THF model and its symmetry shortly. 
 In \cref{sec:irrelevance}, based on a poor man's scaling analysis and several experimental facts, we argue that the Kondo screening effect is irrelevant at the charge neutrality point (CNP) of MATBG, and hence the ground state at CNP is the previously identified symmetry-breaking correlated insulator \cite{kang_strong_2019,bultinck_ground_2020,seo_ferromagnetic_2019,vafek_renormalization_2020,TBG4,TBG6}.
Then we derive a simpler effective periodic Anderson model to describe active excitations upon the correlated ground state at CNP (\cref{sec:periodic-Anderson}).
In \cref{sec:single-impurity}, we systematically analyze a single-impurity version of the periodic Anderson model derived in \cref{sec:HSI}.
We first use poor man's scaling to obtain the Kondo temperature ($\tk$) as a function of filling (\cref{sec:scaling-main}) and, then, by applying Wilson's numerical renormalization group (NRG)~\cite{wilson_nrg1_1980,wilson_nrg2_1980,bulla_numerical_2008}, obtain a phase diagram characterized by strong coupling fixed points and various LM fixed points (\cref{sec:NRG}).
The strong coupling phase consists of a Kondo regime and a frozen impurity regime, and the gapless states at $|\nu| \gtrsim 1$ are found to be in the Kondo regime. 
The spin susceptibility and the entropy obtained in \cref{sec:LM} further explain the so-called Pomeranchuk effect observed in MATBG \cite{rozen_entropic_2021,saito_isospin_2021}. 
For example, the spin susceptibility obeys Curie's law at high temperatures, suggesting the existence of LM, and approaches a constant at lower temperatures, suggesting a Fermi liquid phase. 
However, in sharp contrast to the actual Pomeranchuk effect in helium, which is a first-order liquid-to-solid phase transition on heating, our theory predicts that the transition from the Fermi liquid to the LM state is a {\it continuous crossover}.

We perform a combined dynamical mean-field theory (DMFT) and Hartree-Fock (HF) calculation to the effective periodic Anderson model in \cref{sec:DMFT}, where the impurity solver is implemented using NRG. Flavor symmetries among the remaining active excitations are assumed in the calculation for simplicity. At non-integer fillings and $|\nu|=3$, the DMFT+HF calculation predicts heavy Fermi liquid when the $f$ orbitals are partially filled (\cref{sec:DMFT-spectral}). At the integer fillings $|\nu|=1,2$, the ground state is either a symmetric Mott insulator or a heavy Fermi liquid with extremely low $\tk$, depending on the parameters of the Hamiltonian. The calculation also reproduces the zero-energy peak ~\cite{nuckolls_strongly_2020,choi_correlation-driven_2021,choi_interaction-driven_2021,oh_evidence_2021} and the transition cascade~\cite{wong_cascade_2020,choi_interaction-driven_2021} seen in the STM spectrum, and the saw-tooth inverse compressibility \cite{rozen_entropic_2021,wong_cascade_2020,zondiner_cascade_2020,park_flavour_2021,saito_isospin_2021}. We also compare the DMFT+HF results to the single impurity results and find that the single impurity model yields reasonable estimations of $\tk$ and the entropy but misses the possible Mott insulators. In \cref{sec:DMFT-competition}, we discuss the competition between the Kondo screening and the RKKY interaction. We show that the RKKY interaction dominates near fillings $|\nu|=1,2$, leading to symmetry-breaking states at zero temperature. In \cref{sec:DMFT-bands}, we explicitly calculate the heavy Fermi liquid bands and Fermi surfaces at $T\ll\tk$ and $T>\tk$. A {\it smoking gun} signature of the heavy Fermi liquid is the expansion of Fermi surface on cooling. This signature can be used to verify our theory in future experimental studies. Finally, we briefly summarize this work and discuss its possible relevance to superconductivity in \cref{sec:discussions}.

\section{The effective model}

\subsection{Topological heavy fermion model}
\label{sec:THF}

\begin{figure}
\centering
\includegraphics[width=\linewidth]{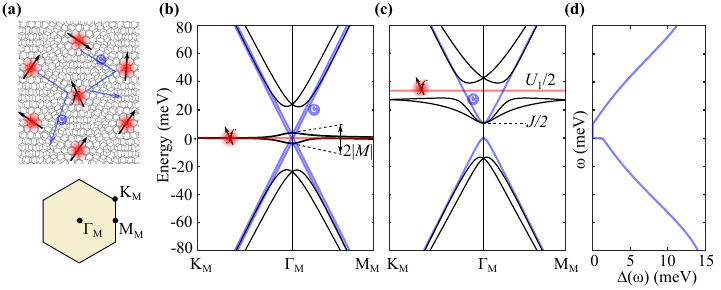}
\caption{The THF model.
(a) Top: red spheres represent the effective $f$-electrons located at AA-stacking regions of MATBG, and blue spheres represent the itinerant $c$-electrons. Bottom: the moir\'e Brillouin zone.  
(b) Black bands are given by the free part of the THF model ($\hH_0$ in \cref{eq:H02v}). 
Red and blue bands are the decoupled $f$- and $c$-bands, respectively.
$M$ is a parameter that determines the bandwidth of the flat bands. 
We focus on the $M\to 0$ limit in this work. 
(c) Band structure of the {\it active} electron modes upon the symmetry-breaking parent state for $\nu>0$, described by \cref{eq:H01v}.
Red and blue bands are the decoupled $f$- and $c$-bands, respectively.
(d) The hybridization function $\Delta(\omega)$ in the single impurity model contributed by the $c$-bands in (c). 
}
\label{fig:model-main}
\end{figure}

One theoretical challenge in studying correlation physics in MATBG is the lack of a fully symmetric lattice model for low energy physics, which is forbidden by the band topology protected by a $C_{2z}T$ symmetry \cite{song_all_2019,po_faithful_2019,ahn_failure_2019} and an emergent particle-hole symmetry $P$ \cite{TBG2} - even though extended Hubbard models \cite{kang_symmetry_2018,koshino_maximally_2018,yuan_model_2018,zou_band_2018,zang_real_2022} can be constructed at the sacrifice of either symmetry or locality. 
The band topology was thought as fragile \cite{song_all_2019,po_faithful_2019,ahn_failure_2019} but was later shown to be a stable symmetry anomaly jointly protected by $C_{2z}T$ and $P$ \cite{TBG2}. 
The THF model \cite{song_magic-angle_2022,shi2022heavy} resolved this problem by ascribing the strong correlation to effective $f$-orbitals at the AA-stacking regions, which form a triangular lattice, and leaving the remaining low energy states to continuous $c$-bands described by a topological Dirac Hamiltonian (\cref{fig:model-main}). 
Its free part is given by 
{\small
\begin{align}
 & \hat{H}_0 = -\mu \hat{N} + \sum_{\eta s}\sum_{aa'}\sum_{|\kk|<\Lambda_c} H^{(c,\eta)}_{aa'}(\kk) c^\dagger_{\kk a\eta s}c_{\kk a\eta s} \nonumber\\
& + \sum_{\eta s \alpha a} \sum_{|\kk|<\Lambda_c} \pare{e^{-\frac{|\kk|^2\lambda^2}2} H^{(cf,\eta)}_{a\alpha}(\kk)c^\dagger_{\kk a\eta s}f_{\kk\alpha\eta s}+h.c.}\ .\label{eq:H02v}
\end{align}}%%
Here $\mu$ is the chemical potential, $\hat{N}$ is the particle-number operator, $c_{\kk a\eta s}$ is the fermion operator for the $c$-electron of the momentum $\kk$, orbital $a$ ($=1,2,3,4$), valley $\eta$ ($=\pm$), and spin $s$ ($=\uparrow,\downarrow$), $f_{\kk \alpha \eta s}$ is the corresponding fermion operator for the $f$-electron of the orbital $\alpha$ (=1,2). 
The momentum of $c$-bands is in principle limited within the cutoff $\Lambda_c$, but the theory yields the same low energy physics in the $\Lambda_c\to \infty$ limit. 
Hence, hereafter we will drop the restriction $|\kk|<\Lambda_c$. 
$H^{(c,\eta)}(\kk) = v_\star (\eta \sigma_x \otimes \sigma_0 k_x - \sigma_y \otimes \sigma_z k_y) + \mathbb{0}_{2\times2}\oplus M \sigma_x$ is the Dirac Hamiltonian of the $c$-bands. 
When $M\neq 0$, $c$-bands have a quadratic band touching at the zero energy, whereas when $M=0$, $c$-bands become linear. 
The two-by-two block of $H^{(cf,\eta)}_{a\alpha}(\kk)$ for $a=1,2$ is given by $\gamma\sigma_0 + v_\star' (\eta \sigma_x k_x + \sigma_y k_y)$, and the two-by-two block of $H^{(cf,\eta)}_{a\alpha}(\kk)$ for $a=3,4$ vanishes. 
The parameter $\lambda$ in the second line of \cref{eq:H02v} is the spread of the Wannier functions of $f$-electrons, and it truncates the hybridization at $|\kk| \gg \lambda^{-1}$. 
In this work we adopt the parameters of Ref.~\cite{song_magic-angle_2022}: $\gamma=-24.75$meV, $v_\star=-4.303\mathrm{eV\cdot\mathring{A}}$, $v_\star'= 1.623 \mathrm{eV\cdot\mathring{A}}$, $\lambda = 1.413 1/k_\theta$, $k_\theta =1.703\mathring{\rm A}^{-1} \cdot 2\sin \frac{\theta_m}2$ with $\theta_m=1.05^\circ$ being the first magic angle. 
The resulting band structure with a nonzero $M$ (3.697meV) is shown in \cref{fig:model-main}(b). 
One can see that the topological flat bands result from the hybridization between $c$- and $f$-bands and have a bandwidth $2|M|$.

In each valley $\eta$, the Hamiltonian $\hH_0$ respects a magnetic space group $P6'2'2$ \cite{song_all_2019} (\#177.151 in the BNS setting \cite{BCS-Mag}), generated by $C_{3z}=e^{i\eta \frac{2\pi}3 \sigma_z} \oplus e^{i\eta \frac{2\pi}3 \sigma_z} \oplus \sigma_0$, $C_{2x}=\mathbb{1}_{3\times 3} \otimes \sigma_x$, and $C_{2z}T=\mathbb{1}_{3\times 3} \otimes \sigma_x K$ (with $K$ being the complex conjugation), and translations along $\mathbf{a}_{1,2}= \frac{2\pi}{3k_\theta}(\pm\sqrt{3},1)$. 
The first (second) two-by-two (four-by-four) block in the operators act on the $f$-electrons ($c$-electrons). 
The interaction Hamiltonian given in the following paragraph also respects these crystalline symmetries.

The interaction Hamiltonian is given by
\begin{widetext}
{\small
\begin{align}
\hH_I= & \frac{U_1}{2}\sum_{\RR} \delta n^f_{\RR} \delta n^f_{\RR} 
    +\frac{U_2}{2}\sum_{\langle \RR \RR'\rangle} \delta n^f_{\RR} \delta n^f_{\RR'} 
    +\frac{1}{2N_M}\sum_{\qq aa'}V(\textbf{q}) \delta n_{-\qq a'}^{c} \delta n^c_{\qq a}
    +\frac{1}{N_M}\sum_{\RR \qq a} W_a e^{-i\qq\cdot\RR} \delta n^f_{\RR} \delta n^c_{\qq a}
    -\frac{J}{2N_M} \sum_{\substack{ \eta\eta'\alpha\alpha'\\ ss'}}\sum_{\substack{\kk,\kk' \\ \RR}} \bigg[
    \nonumber\\
& (\eta\eta'+(-1)^{\alpha+\alpha'}) e^{-i(\kk-\kk')\cdot\RR}
(f^\dagger_{\RR\alpha'\eta' s'}f_{\RR\alpha\eta s} - \frac12\delta_{\eta\eta'}\delta_{\alpha\alpha'} \delta_{ss'} )
 (c^\dagger_{\kk ,\alpha+2\eta s}c_{\kk',\alpha'+2\eta's'} -\frac12 \delta_{\kk\kk'}\delta_{\eta\eta'}\delta_{\alpha\alpha'} \delta_{ss'}) \bigg] \ , \label{eq:HI2v}
\end{align}}%%
\end{widetext}
where $N_M$ is the number of moir\'e cells, 
$f_{\RR \alpha\eta s}$ is the real space fermion operator for the $f$-electrons, 
$\RR$'s form the triangular lattice shown in \cref{fig:model-main}(a), $\inn{\RR\RR'}$ represents nearest neighbor pairs (ordered),
$\delta n^f_{\RR} = \sum_{\alpha\eta s} (f^\dagger_{\RR\alpha\eta s}f_{\RR\alpha\eta s} -\frac12)$ is the total density operator (counted from CNP) of $f$-electrons at $\RR$, 
$\delta n^c_{\qq a} = \sum_{\eta s \kk} (c^\dagger_{\kk+\qq a \eta s} c_{\kk a\eta s}-\frac12\delta_{\qq0})$ is the density operator for $c$-electrons of the orbital $a$.  
$U_{1,2}$, $V(\qq)$, $W_{a}$ are the density-density interaction between $ff$, $cc$, $cf$ electrons, respectively, and $J$ is an exchange interaction between $cf$ electrons.
We adopt the parameters $U_1=57.95$meV, $U_2=1.16$meV, $W_1=W_2=44.03$meV, $W_3=W_4=50.20$meV, $J=16.38$meV, and $V(\qq) =V_0 \frac{\tanh(\xi|\qq|/2)}{\xi|\qq|/2}$, with  $\xi=10$nm and $V_0 = 48.33$meV \cite{song_magic-angle_2022}. 
As explained in \cref{app:U2}, the value of $U_2$ is suppressed from the original value in Ref.~\cite{song_magic-angle_2022}. 

Hereafter, we mainly focus on the flat-band limit where $M=0$, which has been shown as a good approximation using realistic parameters \cite{TBG3,TBG4}. 
In this limit, an exact U(4) symmetry of $\hH_0+\hH_I$ between the spin, valley, and orbital flavors emerges, as previously recognized in the projected Coulomb Hamiltonian of the continuous model \cite{kang_strong_2019,bultinck_ground_2020,seo_ferromagnetic_2019, TBG3,vafek_renormalization_2020}.
This U(4) symmetry is {\it not} related to the so-called chiral limit \cite{tarnopolsky_origin_2019,wang_chiral_2021}, which leads to a distinct U(4) symmetry [\onlinecite{TBG3}, \onlinecite{bultinck_ground_2020}]. 
The sixteen U(4) generators acting on  $f_{\RR \alpha \eta s}$, $c_{\kk a\eta s}$ ($a=1,2$), and  $c_{\kk a\eta s}$ ($a=3,4$) are 
\begin{equation}
\Sigma^f_{\mu\nu} = \{\sigma_0 \tau_0 \varsigma_\nu, \sigma_y \tau_x \varsigma_\nu, \sigma_y \tau_y \varsigma_\nu, \sigma_0 \tau_z \varsigma_\nu\}\ ,
\label{eq:U4f}
\end{equation}
\begin{equation}
\Sigma^{c12}_{\mu\nu} = \{ \sigma_0 \tau_0 \varsigma_\nu, \sigma_y \tau_x \varsigma_\nu, \sigma_y \tau_y \varsigma_\nu,  \sigma_0 \tau_z \varsigma_\nu\}\ ,
\label{eq:U4c12}
\end{equation}
and
\begin{equation}
\Sigma^{c34}_{\mu\nu} = \{ \sigma_0 \tau_0 \varsigma_\nu, -\sigma_y \tau_x \varsigma_\nu, - \sigma_y \tau_y \varsigma_\nu,  \sigma_0 \tau_z \varsigma_\nu\}\ ,
\label{eq:U4c34}
\end{equation}
respectively, where $\varsigma_\nu$ ($\nu=0,x,y,z$) are Pauli matrices acting in the spin subspace, $\tau_{\mu}$ ($\mu=0,x,y,z$) are Pauli matrices acting in the valley subspace, and $\sigma_{0,x,y,z}$ are Pauli matrices acting in the orbital subspace.  
With the help of the U(4) symmetry, the $J$ term in \cref{eq:HI2v} can be written as a ferromagnetic coupling between the U(4) LM of $f$-electrons and the U(4) LM of $c$-electrons.
(Readers may refer to the supplementary section S2G of Ref.~\cite{song_magic-angle_2022} for the discussion of the U(4) symmetry, Eqs.~(S106)-(S109) for the definition of U(4) generators, and Eqs.~(S202)-(S209) for why the $J$ term is a U(4) ferromagnetic coupling.) 
When $M\neq 0$, only the $\mu=0,z$ U(4) generators commute with the Hamiltonian, lowering the symmetry group to U(2)$\times$U(2). 
The rotation generated by $\mu=z$, $\nu=0$ is referred to as the valley-U(1) symmetry.

Consistent with previous results \cite{kang_strong_2019,bultinck_ground_2020,seo_ferromagnetic_2019,vafek_renormalization_2020,TBG4,TBG6}, a Hartree-Fock treatment of the THF model has predicted the ground state at CNP with $M=0$ as a U(4) LM state lying in a 20-fold multiplet that corresponds to the $[2,2]$ representation of U(4) group \cite{TBG4}. Each degenerate state in the manifold respects a U(2)$\times$U(2) subgroup \cite{song_magic-angle_2022}.
These states can be approximately written as
{\small
\begin{equation}
\ket{\Psi_0}= e^{-i\theta_{\mu\nu} \hat{\Sigma}_{\mu\nu}} \prod_{\RR} f_{\RR1+\up}^\dagger f_{\RR1+\down}^\dagger f_{\RR2+\up}^\dagger 
f_{\RR2+\down}^\dagger |{\rm FS}\rangle \ ,
\label{eq:CNP-state}
\end{equation}}%%
where the $\rm |FS\rangle$ is the Fermi sea state with the half-filled $c$-bands, $\hat{\Sigma}_{\mu\nu}$'s are the U(4) generator operators defined by the matrices in \cref{eq:U4f,eq:U4c12,eq:U4c34}, and $\theta_{\mu\nu}$'s are the rotation parameters. An implicit summation over repeated $\mu,\nu$ indices is assumed. 
When $\theta_{\mu\nu}$'s are zero, $\ket{\Psi_0}$ is the valley-polarized state because all the occupied $f$-electrons are in the $\eta=+$ valley, and the U(2)$\times$U(2) subgroup is generated by $\hat{\Sigma}_{0\nu}$ and $\hat{\Sigma}_{z\nu}$ ($\nu=0,x,y,z$). 
For nonzero $\theta_{\mu\nu}$'s, $\ket{\Psi_0}$ respects an equivalent U(2)$\times$U(2) subgroup. 
In particular, the Kramers inter-valley coherent states can be obtained by setting $\theta_{x0}$ and $\theta_{y0}$ to be nonzero and satisfy $\theta_{x0}^2 + \theta_{y0}^2=(\pi/4)^2$.
When $M\neq0$, the Kramers inter-valley coherent states are found to have a lower energy ($\sim$0.1meV) than the valley polarized states \cite{TBG4,song_magic-angle_2022}.

\subsection{Irrelevance of Kondo screening at CNP}
\label{sec:irrelevance}

Here we argue that the Kondo screening effect is irrelevant at CNP; hence, the U(4) LM state in \cref{eq:CNP-state} is valid as an approximate ground state.
We first examine the energy scale of a fully symmetric Kondo state at CNP. 
Since the $f$-sites are almost decoupled from each other, a reasonable approximation is to view each $f$-site as a single Anderson impurity coupled to a bath of $c$-electrons. 
If we only consider the on-site $U_1$ interaction and the hybridization between $f$- and $c$-electrons ($H^{(cf,\eta)}(\kk)$ in \cref{eq:H02v}), then it is almost a standard Anderson model with eight flavors.
The effect of $c$-bath is described by the hybridization function $\Delta(\omega)$, defined as the imaginary part of the (retarded) self-energy of a free $f$-electron (in the absence of $U_1$) coupled to the $c$-bath, \ie $\Im[\Sigma^{0}_{\alpha\eta s, \alpha'\eta' s'} (\omega)] = -\delta_{\alpha,\alpha'} \delta_{\eta\eta'} \delta_{ss'} \Delta(\omega)$. 
The identity matrix form of $\Im[\Sigma_{\alpha\eta s, \alpha'\eta' s'} (\omega)]$ is guaranteed by the spin-SU(2) ($\delta_{ss'}$), the valley-U(1) and the time-reversal ($\delta_{\eta\eta'}$), and crystalline ($\delta_{\alpha\alpha'}$) symmetries. 
In the flat-band limit ($M=0$), the linear dispersion of $c$-bands (\cref{fig:model-main}(b)) leads to a linear-in-energy density of states as well as a linear-in-energy hybridization function, \ie $\Delta(\omega)\sim |\omega|$.
As a consequence, low-lying states of the impurity will see vanishing bath electrons when the energy scale is small enough. 
Both numerical \cite{chen_kondo_1995,gonzalez-buxton_renormalization-group_1998,ingersent_critical_2002} and analytical \cite{fritz_phase_2004} RG studies have shown that Anderson impurity models with such a $\Delta(\omega)\sim |\omega|$ hybridization function do {\it not} have the strong coupling fixed point that exhibits Kondo screening. 
Instead, the only stable fixed point is the LM phase. 

With a finite $M$, the $c$-bands given by $H^{(c,\eta)}(\kk)$ in \cref{eq:H02v} have a quadratic band touching at the zero energy, \ie $\pm (-M/2+\sqrt{M^2/4 + v_\star^2 \kk^2})$, leading to a finite $\Delta(0)$. 
Nevertheless, the Kondo energy scale resulting from realistic parameters is still negligible.
In \cref{app:RG-CNP} we derived an analytical expression of $\Delta(\omega)$ for the symmetric state at CNP.
In the low energy regime ($|\omega|<U_1/2$), we find $\Delta(\omega) \approx 2 \Delta(0)  |\omega|/M$ for $|\omega|>M$ and $\Delta(\omega) \approx \Delta(0) ( 1 + |\omega|/M)$ for $|\omega|<M$. 
To estimate the Kondo energy scale, we apply a poor man's scaling (detailed in \cref{app:RG-CNP}) that considers the $\omega$-dependence of $\Delta(\omega)$.
There are two stages in the RG process: 
(i) energy scale from $U_1/2$, above which the perturbation theory no more applies, falls down to $M$, 
(ii) energy scale gets renormalized below $M$. 
RG in the first stage effectively enhances $\Delta(0)$ to  $g_1 \Delta(0)$ with $g_1>1$ a factor determined by $M$. 
Then, RG in the second stage gives the Kondo energy scale
\begin{align} \label{eq:Kondo-scale-CNP}
D_K  \approx M e^{ 1 - \frac{\pi U_1}{4 \NN g_1 \Delta(0)} } 
\end{align}
where the factor $e^1$ is contributed by the linear $\omega$-dependence of $\Delta(\omega)$ in the second stage, and $\mathcal{N}=8$ is the number of flavors. 
$D_K$ strongly depends on the actual bandwidth. 
For $2M=$5, 7.4, 10, 15, 20meV and fixed $U_1=57.95\mathrm{meV}$, we obtain $D_K\approx5.1\times 10^{-6}$, $3.8\times10^{-4}$, $4.3\times 10^{-3}$, $4.9\times10^{-2}$, and 0.17meV, respectively.
For $U_1=$60, 50, 40, 30meV and fixed $2M=10$meV we have $D_K\approx 3.5\times10^{-3},1.0\times 10^{-2},2.8\times10^{-2},8.0\times10^{-2}$meV, respectively.
Given that the single-particle bandwidth estimated by the BM model and the first-principle calculations are about 7.4meV \cite{song_magic-angle_2022} and 10meV \cite{lucignano_crucial_2019,carr_exact_2019}, respectively, $D_K$ should be at most at the order of $10^{-2}$meV. 
This energy scale is still much lower than the energy gain of the symmetry-breaking correlated state \cite{TBG4,TBG5}.  
The bandwidth of the Goldstone modes at CNP from $\Gamma_M$ to $M_M$ is about 8meV. (See Fig.~2 of Ref.~\cite{TBG5}). 
If we understand this spectrum as a tight-binding band of the Holstein–Primakoff bosons on the $f$-sites, which form a triangular lattice, then the nearest neighbor hopping is about 8meV/8=1meV.
This hopping indicates an RKKY interaction much larger than the Kondo energy scale. 
(In twisted bilayer graphene at non-magic angles, the bandwidth $2M$ can be much larger, and a symmetric Kondo phase could be stabilized at CNP if the $f$ orbitals still remain well localized.)

In addition, as we have neglected all the interactions except $U_1$ in the estimation, the single-impurity model has a U(8) symmetry.
A U(8) breaking must be caused by other interaction terms, \eg $J$ in \cref{eq:HI2v}, and will lead to a multiplet splitting. 
When the energy scale in the RG is smaller than the multiplet splitting, the degeneracy factor $\NN$ should be reduced accordingly, and $D_K$ will be further suppressed \cite{coleman_introduction_2015}. 
Therefore, we conclude that the ground state of MATBG at CNP is a symmetry-breaking state.

The symmetry-breaking state at CNP is also supported by various experiments. 
In contrast to the Kondo resonance, STM measurements have shown strong suppression of the density of states at the zero energy at CNP \cite{xie_spectroscopic_2019,wong_cascade_2020,nuckolls_strongly_2020,choi_interaction-driven_2021,choi_correlation-driven_2021,oh_evidence_2021,choi_electronic_2019,jiang_charge_2019,kerelsky_maximized_2019}. 
Some transport experiments \cite{lu_superconductors_2019,polshyn_large_2019,stepanov_untying_2020,saito_independent_2020} also exhibit a gap behavior at CNP. 
Although there are also transport experiments showing semimetal behavior, the gaplessness can be explained if there are fluctuations of the local moments from site to site, which is possible due to the Goldstone mode fluctuations \cite{TBG5,khalaf2020soft} and possible inhomogeneity of the sample.

\subsection{Effective periodic Anderson model for \texorpdfstring{$\nu>0$}{nu>0} states}
\label{sec:periodic-Anderson}

We aim for an effective model describing the {\it active} excitations upon the ground state $\ket{\Psi_0}$ (\cref{eq:CNP-state}) at CNP.
Let us first assume the valley-polarized state, where $\theta_{\mu\nu}$'s in \cref{eq:CNP-state} are all zero such that all the occupied $f$-electrons are in the $\eta=+$ valley. 
As detailed in the supplementary material of Ref.~\cite{song_magic-angle_2022} and in Ref.~\cite{TBG5}, the lowest electron and hole excitations are in the $\eta=-$ and $\eta=+$ valleys, respectively. 
Thus, for a small electron doping, only excitations in the $\eta=-$ valley will be involved, and the electrons in the $\eta=+$ valley can be viewed as a static background.
The effective Hamiltonian can be obtained by replacing operators in the $\eta=+$ valley by their expectation values on $\ket{\Psi_0}$, which are 
$\langle f^\dagger_{\RR\alpha+ s}f_{\RR'\alpha'+ s'}\rangle=\delta_{\RR\RR'}\delta_{\alpha\alpha'}\delta_{ss'}$, 
$\inn{c^\dagger_{\kk a +s }c_{\kk' a' +s' }}\approx \frac{1}{2}\delta_{\kk \kk'}\delta_{aa'}\delta_{ss'}$,
$\inn{c^\dagger_{\kk a + s}f_{\RR \alpha + s'}}=0$.
Substituting these expectation values into $\hH_0+\hH_I$, we obtain the effective free Hamiltonian 
{\small
\begin{align} \label{eq:H01v}
& \hat{H}_{0}^{\rm eff} = -\mu \hat{N} + \sum_{{\kk s}aa'} \pare{  H_{a a'}^{(c)}(\kk)   + \frac{J}2 \delta_{aa'} (\delta_{a3}+\delta_{a4})  }c_{\kk a s}^\dagger c_{\kk a' s} 
\nonumber \\
& +   \frac{U_1}2 \sum_{\RR } n^f_{\RR} + \sum_{\substack{\kk a\alpha s}} \pare{ e^{-\frac12\lambda^2\kk^2} H^{(cf)}_{a \alpha}(\kk) c^\dagger_{\kk a s}f_{\kk \alpha s} + h.c.} ,
\end{align}}%%
where $n_{\RR\alpha s} = \sum_{\alpha s} f_{\RR \alpha s}^\dagger f_{\RR \alpha s}$ is the density operator of $f$-electrons at $\RR$.
Here we have dropped the valley index $\eta$ as they are limited to $\eta=-$. 
The $H^{(c)}(\kk)$ and $H^{(cf)}(\kk)$ matrices are given by the $H^{(c,-)}(\kk)$ and $H^{(cf,-)}(\kk)$ matrices defined after \cref{eq:H02v}. 
The effective interaction Hamiltonian is
{\scriptsize
\begin{align}\label{eq:HI1v}
& \hH^{\rm eff}_I  = \frac{U_1}{2}\sum_{\RR} :n^f_{\RR}n^f_{\RR}: + \frac{U_2}{2}\sum_{\langle \RR \RR'\rangle}n^f_{\RR}n^f_{\RR'} 
\nonumber\\
& + \frac{1}{2N_M}\sum_{\qq a a'}V(\qq)\delta n^c_{-\qq, a'} \delta n^c_{\qq,a}  + \frac{1}{N_M}\sum_{\RR \qq a} W_a e^{-i\qq\cdot\RR} n^f_\RR \delta n^c_{\qq a} \nonumber\\
& - \frac{J}{N_M} \sum_{\RR ss'}\sum_{\kk \kk' \alpha }e^{-i(\kk-\kk')\cdot \RR }    f^\dagger_{\RR\alpha s'}f_{\RR\alpha s}
 (c^\dagger_{\kk ,\alpha+2,s}c_{\kk',\alpha+2,s'} - \frac12 \delta_{\kk\kk'}\delta_{ss'})\ ,
\end{align}
}%%
where $\delta n^c_{\qq a} = \sum_{s \kk} (c_{\kk+\qq a s}^\dagger c_{\kk a s}-\frac12\delta_{\qq0})$. 
The $U_1$ term in $\hH^{\rm eff}_I$ is normal ordered, and the bilinear term left over in normal-ordering, {\it i.e.}, $\frac{U_1}2 n_{\RR}^f$, is now in $\hH^{\rm eff}_0$. 
The bilinear term in $c$ operators contributed by the $J$ interaction is also in $\hH^{\rm eff}_0$. 
Band structure of $\hH^{\rm eff}_0$ is shown in \cref{fig:model-main}(c).

In the flat-band limit ($M=0$), $\hH_0^{\rm eff} + \hH_I^{\rm eff}$ also applies to arbitrary U(4) partners of the valley-polarized state, including the so-called Kramers intervalley coherent state. 
To be specific, for a generic $\ket{\Psi_0}$ given in \cref{eq:CNP-state}, we can always define rotated operators $f_{\RR \alpha s} = U f_{\RR \alpha - s} U^\dagger $, $c_{\kk a s} =  U c_{\kk a - s} U^\dagger$, where $U=e^{-i \theta_{\mu\nu}\hat{\Sigma}_{\mu\nu}}$ is the U(4) rotation defining $\ket{\Psi_0}$, such that the effective Hamiltonian on the rotated basis is the same as \cref{eq:H01v,eq:HI1v}. The effective Hamiltonian $\hH_0^{\rm eff} + \hH_I^{\rm eff}$ respects all the crystalline symmetries discussed in \cref{sec:THF}. 
In the flat-band limit ($M=0$), $\ket{\Psi_0}$ respects a U(2)$\times$U(2) subgroup of the U(4) group, \eg independent spin-charge rotations in the two valleys for the valley polarized $\ket{\Psi_0}$. 
However, since the effective Hamiltonian only involves half of the d.o.f., \eg the active $\eta=-$ valley for the valley polarized $\ket{\Psi_0}$, only one U(2) factor is meaningful for $\hH_0^{\rm eff} + \hH_I^{\rm eff}$. 
Therefore, hereafter we will say that $\hH_0^{\rm eff} + \hH_I^{\rm eff}$ respects a U(2) symmetry group. 

\begin{table}[t]
\centering
\begin{tabular}{c|c|c|c|c}
\hline
\hline
    &  $\hH_0+\hH_I$ & $\hH_0^{\rm eff} + \hH_I^{\rm eff}$ & $\hH_{\rm SI}$ ($J_H=0$) & $\hH_{\rm SI}$ ($J_H>0$) \\
\hline
$M=0$   & U(4) & U(2) & U(4) & U(2)$\times$U(2)  \\
$M\neq 0$  & U(2)$\times$U(2) & U(2) & U(4) & U(2)$\times$U(2) \\
\hline
\hline
\end{tabular}
\caption{Continuous symmetries of the effective models. 
$\hH_0+\hH_I$ is the original THF model.
For $\nu>0$ ($\nu<0$),  $\hH_0^{\rm eff} + \hH_I^{\rm eff}$ is the effective periodic Anderson model for the active particle (hole) excitations upon the symmetry-breaking state at CNP. 
$\hH_{\rm SI}$ is a single-impurity version of $\hH_0^{\rm eff} + \hH_I^{\rm eff}$.
$J_H$, estimated as $0.3$meV, is an effective Hund's coupling of $f$-electrons resulted from the exchange coupling ($J$) between $c$- and $f$-electrons. 
}
\label{tab:symmetry}
\end{table}

As discussed at the end of \cref{sec:THF}, when $M\neq 0$, the U(4) symmetry is broken, and the ground state is the Kramers intervalley coherent state. 
As a consequence, the effective Hamiltonian should have additional terms.
However, $M$ will not further lower the crystalline and U(2) symmetries of $\hH^{\rm eff}_0 + \hH^{\rm eff}_I$ and hence will only play a minor role in the effective theory.
In \cref{app:M} we treat $M$ perturbatively in terms of $M/J$, which is about $0.2$ using the parameters given in \cref{sec:THF}.
We find that the leading order correction is simply an energy shift of the $c$-electrons.
To avoid confusion, in \cref{tab:symmetry} we summarize the continuous symmetries of different Hamiltonians discussed in this work. 

The effective model for $\nu<0$ states, which only involves hole excitations, can be obtained by applying the particle-hole operation $\mathcal{P}_c$ \cite{TBG3,song_magic-angle_2022} to $\hH_0^{\rm eff} + \hH_I^{\rm eff}$. 

\section{Solution to the single-impurity model}
\label{sec:single-impurity}

In this section, we focus on a single-impurity version of $\hH^{\rm eff}_0 + \hH^{\rm eff}_I$, where only the correlation effects at the $\RR=0$ $f$-site are considered. 
Interactions not involving this $f$-site will be treated at the mean-field level. 
The single-impurity model successfully explains a number of experimental features in the metallic phases despite symmetry-breaking gaps at integer fillings and indicates that the metallic phases are heavy Fermi liquids.
For a complete discussion including the possible symmetry breaking at integer fillings, we leave it to the periodic Anderson model investigated in \cref{sec:DMFT}.

\subsection{Single impurity model for \texorpdfstring{$\nu>0$}{nu>0} states}
\label{sec:HSI}

At a given filling $\nu$, the mean-fields are characterized by only a few parameters: $\nu_f = \inn{n^f_{\RR}} $, $\nu_{c,a} = \frac1{N_M} \inn{\delta n^c_{\qq=0, a} }$, where $\nu_{c,1}=\nu_{c,2}$, $\nu_{c,3}=\nu_{c,4}$ due to the crystalline symmetries. 
The considered correlated site at $\RR=0$ is described by the Hamiltonian 
\begin{equation} \label{eq:Hf}
\hat{H}_f = \epsilon_f n^f + \frac{U_1}2  :n^f n^f: \ ,
\end{equation}
where the lattice index $\RR$ ($=0$) is omitted for simplicity, $\epsilon_f =  6\nu_fU_2 + \sum_a\nu_{c,a}W_a + \frac{1}{2}U_1 - \frac12 J \nu_{c,3} - \mu$ is the mean field level of the $f$-site. 
The $U_2$, $W_a$, $J$ terms in $\epsilon_f$ are contributed by the Hartree mean-fields of interactions in \cref{eq:HI1v}. 
The effective Hamiltonian of $c$-electrons is given by
\begin{equation} \label{eq:Hc}
\hH_c =  \sum_{\kk s aa'} 
 \brak{ H^{(c)}_{aa'}(\kk) + \delta_{aa'}\epsilon_{c,a}}c_{\kk a s}^\dagger c_{\kk a' s} \ ,
\end{equation}
where $H^{(c)}(\kk)$ is the free Dirac Hamiltonian in \cref{eq:H01v}, and $\epsilon_{c,1} = \epsilon_{c,2} =\nu_fW_1 + \nu_c V_0 - \mu$, $\epsilon_{c,3}=\epsilon_{c,4}= \nu_f (W_3  - \frac{J}4) + \nu_c V_0 +\frac{J}2 - \mu$ are the mean-field levels of $c$-electrons. 
$\epsilon_{c,1}$ is contributed by Hartree mean-fields of $W_1$ and $V$ interactions in \cref{eq:HI1v} and $\epsilon_{c,3}$ is contributed by the Hartree mean-fields of $W_3,V$ and $J$ interactions in \cref{eq:HI1v}.
The band structure of \cref{eq:Hc} is given by  $(\eco+\ect)/2 \pm \sqrt{ G^2/4 + v_\star^2 \kk^2 }$, where $G=\epsilon_{c,3}-\epsilon_{c,1}$ is the band gap.
$c$-bands with $\epsilon_{c,1}=0$, $\epsilon_{c,3}=J/2$ are shown in \cref{fig:model-main}(c).
Since the interaction $V(\qq)$ of $c$-electrons is completely treated at the mean-field level, $\hH_c$ is an effective free-fermion system.

The $f$-site is coupled to $c$-electrons via the $H^{(cf)}$ term in \cref{eq:H01v} and $H_J$ in \cref{eq:HI1v}.  
As detailed in \cref{app:Hund}, these two terms can be treated separately due to $C_{3z}$ symmetry.
The $H_J$ interaction leads to, in addition to the Hartree mean fields discussed in the last paragraph, an effective Hund's coupling of $f$-electrons
\begin{equation}
\hat{H}_H = J_H \sum_{\alpha} f_{\alpha \up}^\dagger f_{\alpha \up} f_{\alpha \down}^\dagger f_{\alpha \down} \ \label{eq:Hund},
\end{equation}
where $J_H$ is about 0.3meV. Since $J_H$ is much smaller than other interactions, we mainly focus on the $J_H=0$ model in the main text and leave discussions for $J_H>0$ to \cref{app:RG-nu123}. 
The $H^{(cf)}$ term leads to a (retarded) self-energy correction $\Sigma^{0}_{\alpha s, \alpha' s'}(\omega)$ to the $f$-electrons, whose imaginary part defines the hybridization function $\Delta (\omega)$, \ie $\Im[\Sigma^{0}_{\alpha s, \alpha' s'}(\omega)] = -\delta_{\alpha \alpha'} \delta_{s s'}\Delta(\omega)$.
The identity matrix structure of the self-energy is guaranteed by SU(2) spin rotation symmetry and crystalline symmetries. 
In \cref{app:hyb} we derived the following analytical expression for $\Delta(\omega)$
{
\begin{align} \label{eq:hybridization-main}
\Delta(\omega)=& \frac{\Omega_0}{4v_\star^2} \abs{\omega - \epsilon_{c,3}} \left(\gamma^2+v_\star^{\prime 2}k_F^2\right)e^{-k_F^2\lambda^2}\nonumber\\
& \times \left[\theta(\omega-\epsilon_{c,3})  +\theta(\epsilon_{c,1}-\omega) \right]\ ,
\end{align}}%%
where $k_F$ is determined by $v_\star^2 k_F^2 + G^2/4 = [\omega - (\eco+\ect)/2]^2$ for either $\omega>\ect$ or $\omega<\eco$. 
As shown in \cref{fig:model-main}(d) (with $\epsilon_{c,1}=0$, $\epsilon_{c,3}=J/2$), $\Delta(\omega)$ has an abnormal $\omega$-dependence compared to those in usual metals.

Baths giving rise to the same $\Delta(\omega)$ are physically equivalent.
We introduce the following effective single-impurity Hamiltonian that yields the same $\Delta(\omega)$ following \cite{bulla_numerical_2008}
\begin{align}\label{eq:HSI}
\hH_{\rm SI} =& \hH_f + \sum_{\alpha s} \int_{-D}^D d \ee \cdot \ee \cdot d_{\alpha s}^\dagger(\ee) d_{\alpha s}(\ee) \nonumber \\
+ & \sum_{\alpha s} \int_{-D}^D d\ee \cdot \sqrt{\frac{\Delta(\ee)}{\pi}} 
    ( f_{\alpha s}^\dagger d_{\alpha s}(\ee) + h.c. )\ , 
\end{align}
where $\hH_f$ is given by \cref{eq:Hf}, and $d_{\alpha s}(\ee)$, satisfying $\{d_{\alpha's'}(\ee'), d_{\alpha s}^\dagger (\ee)\} = \delta_{\alpha'\alpha}\delta_{s's} \delta(\ee'-\ee) $, are the auxiliary bath fermions introduced to reproduce the hybridization function. 
$\hH_{\rm SI}$ is determined by four parameters: $\eef$ the energy level of $f$-electrons, $U_1$ the Coulomb repulsion, $\epsilon_{c,1}$ and $\epsilon_{c,3}$ the energy level  of $a=1,2$ and $a=3,4$ $c$-electrons(or $\eco$ and $G$ equivalently).  
As explained at the beginning of this subsection, the actual values of $\epsilon_{f}$, $\epsilon_{c,a}$ depend on the occupations $\nu_f$, $\nu_{c,a}$.
In this section, we estimate $\nu_f$, $\nu_{c,a}$ by a symmetric self-consistent HF calculation of $\hH^{\rm eff}_0 + \hH^{\rm eff}_I$. 
The obtained $\epsilon_{f}$, $\epsilon_{c,1}$, and $G$ as functions of $\nu$ are shown in \cref{fig:phase}(a). 
A better treatment of these parameters should be a full self-consistent DMFT+HF calculation, which will be carried out in \cref{sec:DMFT}. 
As explained in the following subsections, the essential physics of the Kondo phase is already captured by this single-impurity model with $\nu_f$, $\nu_{c,a}$ estimated by the HF mean-field.

It is worth mentioning that \cref{eq:HSI} has an emergent U(4) symmetry because the four flavors of $f$-electrons are symmetric under permutations. 
It is not surprising that a single-impurity model has a higher symmetry than its lattice version.
$J_H$ lowers the symmetry of the single impurity model to U(2)$\times$U(2), while since it is weak compared to other energy scales, we will mainly focus on the U(4) symmetric model in the main text.

\subsection{Poor man's scaling}
\label{sec:scaling-main}

Before going to numerical calculations, we first apply a poor man's scaling to the single impurity model \cref{eq:HSI} to estimate the Kondo energy scale. 
For now, we can regard $\epsilon_{c,a}, \eef$ as independent parameters. 

We assume that the ground state of the detached impurity has $n_f$ ($=$1, 2, 3) occupied $f$-electrons. 
One should not confuse $n_f$ with $\nu_f$ - the expectation value of $f$-occupation after the impurity is coupled to the bath.
The $f$ electron level $\epsilon_f$ must be in the range $-n_f U_1 <\epsilon_f < -(n_f-1)U_1$. 
We apply a Schrieffer-Wolff transformation to \cref{eq:HSI} to obtain an effective Coqblin–Schrieff model where the local Hilbert space of $f$-electrons is restricted to $n_f$ particles. 
The transformation involves virtual particle and hole excitations, the energies of which are $\Delta E_+ =\eef + n_f U_1$ and $\Delta E_- = -\eef - (n_f-1)U_1$, respectively. 
Adding the two contributions, we have 
{\small
\begin{align}\label{eq:HCS}
& \hH = \sum_{\alpha s} \int_{-D}^D d\ee \cdot \ee \cdot d_{\alpha s}^\dagger(\ee) d_{\alpha s}(\ee) + \frac{4 g}{\pi U_1} \sum_{\alpha \alpha' s s'}  \int_{-D}^D d\ee d\ee' \bigg[ \nonumber\\
& \; \times \sqrt{ {\Delta(\ee) \Delta(\ee') } } ( f_{\alpha s}^\dagger f_{\alpha's'} - x \delta_{\alpha\alpha'}\delta_{ss'} )
 d_{\alpha' s'}^\dagger (\ee') d_{\alpha s} (\ee) \bigg] .
\end{align}}%%
The parameters $g, x$ are given by $ $
\begin{equation} \label{eq:initial}
    g = \frac{U_1}4 \pare{ \frac1{\Delta E_+} + \frac1{\Delta E_-} },\qquad
    x = \frac{\Delta E_-}{U_1}\ . 
\end{equation}
$g$ is a dimensionless parameter characterizing the anti-ferromagnetic coupling strength between the impurity and the bath.  
$x$ appears as a ``charge background'' of the $f$-electrons. 
For $\eef = -(n_f-\frac12)U_1$, there is $g = 1$, $x=\frac12$. 
For a generic $\epsilon_f$ in the range $-n_f U_1 <\epsilon_f < -(n_f-1)U_1$, there are $g\ge 1$ and $0< x<1$. 
Flow equations of $g,x$ are derived in \cref{app:RG-UN,app:RG-nu123}, where the divergence of $g$ indicates the strong coupling fixed point that exhibits the Kondo screening.
We notice that $x$ always flows to $ n_f/4 $, \ie the occupation fraction of $f$-electrons. 

One should be careful about the cutoff $D$ in \cref{eq:HCS}.
First, it must be smaller than $\Delta E_+$ and $\Delta E_-$ for the Schrieffer-Wollf transformation to be valid. 
Second, for analytical convenience, we only keep the positive branch of $\Delta(\omega)$ (\cref{eq:hybridization-main}) at $\omega > \ect$ because when $\nu>0$ the negative branch is far away from the Fermi level. 
Hence, we also require $D<-\ect$. 
We can choose $D = \min(-\ect, \Delta E_+, \Delta E_-)$.

\begin{figure}[ht!]
\centering
\includegraphics[width=\linewidth]{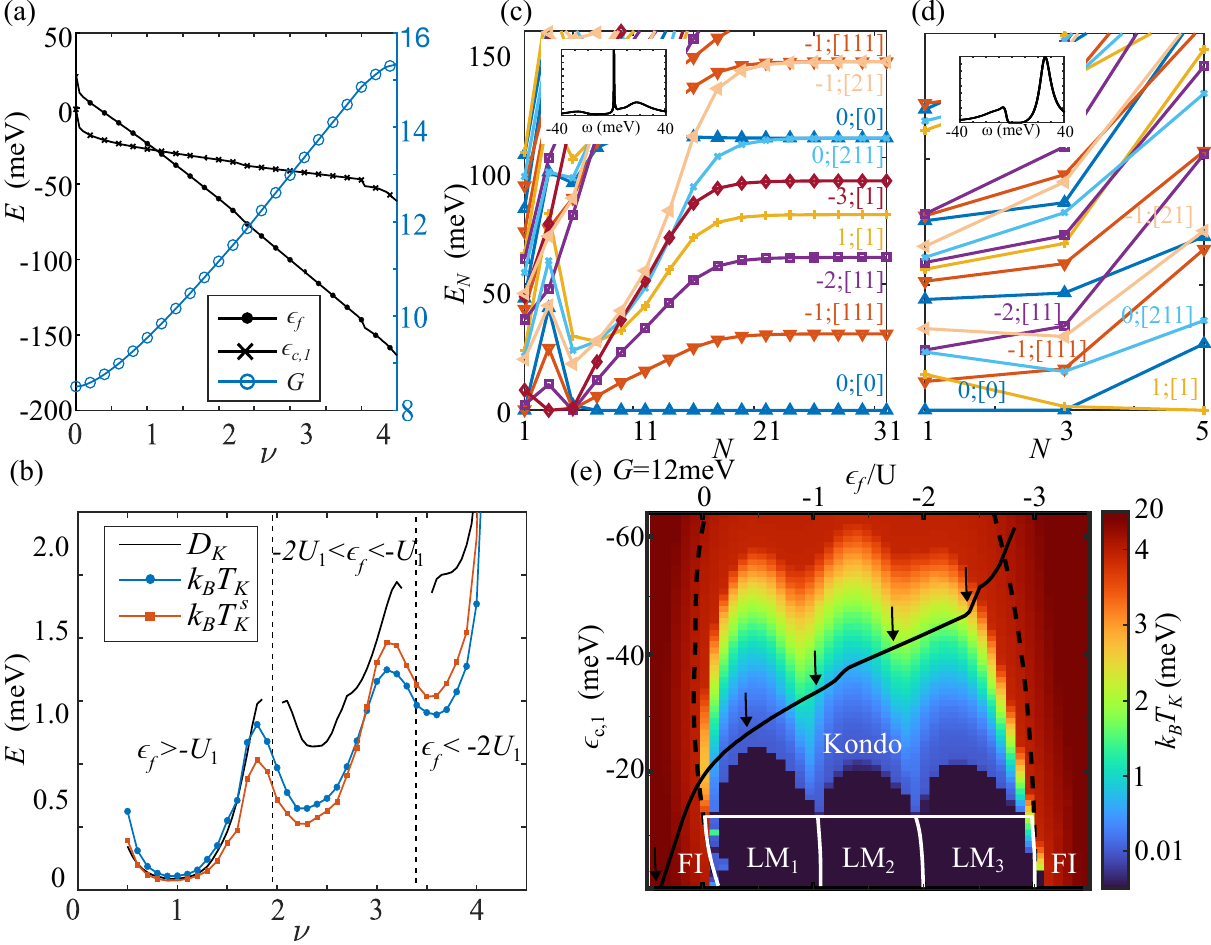}
\caption{Phases and fixed points in the single-impurity model. 
(a) Mean-field values of $\eco,\eef,G$ as functions of the total filling $\nu$. 
(b) The Kondo energy scale estimated by the poor man's scaling ($D_K$), the NRG spectral density ($k_B\tk$), the NRG spin susceptibility ($k_B \tk^s $). 
(c) The RG flows of the many-body spectra of the scaled Hamiltonian $\td{H}_N$ ($N\in$ odd) at $\nu=1.25$ with mean-field parameters $\eco=-28.8$meV, $\eef/U_1=-0.554$, $G=9.92$meV. 
The levels are labeled by total charge $Q$ and the SU(4) irreducible representation. 
The insets are the corresponding spectral densities that exhibit a Kondo resonance. 
(d) The RG flows in the LM$_{1}$ phases, where $\eef=-\frac12U_1$, $\eco=-5$meV, and $G=12$meV.
The quantum numbers of the ground states are emphasized by using larger fonts.
Only spectra at $N<=5$ are shown because the hopping between $N=5$ and $N=6$ bath site in the Wilson chain is as small as $1\times 10^{-19}$meV, which means the bath site at $N\ge 6$ are decoupled from the impurity. 
The insets are the corresponding spectral densities that exhibit local moment features: Hubbard bands without zero-energy peaks.
(e) The phase diagram in the parameter space of $\eco, \eef$ for fixed $G=\ect-\eco$ ($=12$meV). 
The white lines are phase boundaries between local moment (LM) phases and the strong coupling phase.
The dashed black lines are crossover boundaries between the frozen impurity (FI) and Kondo regimes of the strong coupling phase. 
The color maps $\tk$ obtained from NRG spectral density. 
The solid black line indicates the trajectory of $\eco$ and $\eef$ as $\nu$ changes from 0 to 4, where the five arrows from left to right represent $\nu=0,1,2,3,4$, respectively. 
}
\label{fig:phase}
\end{figure}

The flow equation of $g(t)$ as the cutoff is successively reduced to $De^{-t}$  is given by 
\begin{equation}\label{eq:flow1-main}
    \frac{d g} {dt} 
    = \frac{4\Delta(0)}{\pi U_1} \mathcal{N} g^2 + \mathcal{O}( e^{-t} ) \ ,
\end{equation}
and the initial condition $g(0)$ is given by \cref{eq:initial}. 
Here $\NN=4$ is the number of flavors. 
The local Hilbert space for $n_f=1,2,3$ is four-, six-, and four-fold, respectively.  
The $\mathcal{O}(e^{-t} )$ terms originate from particle-hole asymmetry and are irrelevant at small energy scales but they may affect the coupling constant at an early stage of the RG process.
As shown in \cref{fig:model-main}(d), the positive branch of $\Delta(\omega)$ can be well approximated by a linear function, \ie $\Delta(\omega) \approx \Delta(0) ( 1 - \omega/\ect) $.
Using this linear approximation we obtain the Kondo energy scale (\cref{app:RG-nu123})
\begin{equation} \label{eq:TK1-main}
    D_K = D \exp\pare{ y - \frac{\pi U_1}{4 \NN \Delta(0) g(0) } }
\end{equation}
where $y \approx (\frac{\Delta E_+}{U_1} + \frac12 - \frac12 n_f) \frac{D}{\ect}$ is factor contributed by the irrelevant $\mathcal{O}(e^{-t})$ terms at $\NN=4$.
Noticing $\ect<0$, at fixed $\Delta E_+$ a smaller $n_f$ means smaller $y$ and suppresses the Kondo energy scale as this means that virtual processes contributing to the RG equation involve more hole excitations in the bath, which has a smaller $\Delta(\omega)$. 
In \cref{fig:phase}(b) we plot the obtained $D_K$ as a function of the filling $\nu$ using the mean-field $\epsilon_{c,a},\eef$ parameters given in \cref{fig:phase}(a).
The estimated Kondo energy scale is at the order of 1meV.

As explained in the end of \cref{sec:HSI}, if the Hund's coupling in \cref{eq:Hund} is considered, the U(4) symmetry of $\hH_{\rm SI}$ will be reduced to U(2)$\times$U(2). 
Then the six-dimensional local Hilbert space in the $n_f=2$ case will split: the four states with $(n^f_{1\uparrow},n^f_{1\downarrow}; n^f_{2\uparrow}, n^f_{2\downarrow})=$(10;10), (10;01), (01;10), (01;01) do not feel $J_H$ and have the energy $2\eef + U_1$, whereas the two states (11;00), (00;11) have the energy $2\eef + U_1 + J_H$. 
As explained in detail in \cref{app:RG-nu123}, this multiplet splitting will further suppress the Kondo energy scale if $D_K$ given by \cref{eq:TK1-main} is smaller than $J_H$.
However, as shown in \cref{fig:phase}(b), $D_K$ given by \cref{eq:TK1-main} for $n_f=2$ is always larger than $J_H$, which is estimated as $0.3$meV. 
Hence the multiplet splitting plays a minor role. 
This argument further justifies our approximation of neglecting $J_H$.

\subsection{NRG phase diagram}
\label{sec:NRG}

We now apply the NRG approach \cite{wilson_nrg1_1980,wilson_nrg2_1980,bulla_numerical_2008} to study the single-impurity model $\hH_{\rm SI}$.
In this approach, the bath is alternatively realized by a Wilson chain 
{\small
\begin{align} \label{eq:HN-maintext}
& \hat{H}_N = \hat{H}_{f} + \sum_{\alpha s} t_0(f_{\alpha s}^\dagger d_{1\alpha s} + h.c.) \nonumber \\
& + \sum_{n=1}^{N} \sum_{\alpha s} \epsilon_n d_{n\alpha s}^\dagger d_{n\alpha s}  
+ \sum_{n=1}^{N-1} \sum_{\alpha s} (t_n d_{n+1\alpha s}^\dagger d_{n\alpha s} + h.c.)\ ,
\end{align}}%%%
where $N$ is the length of the Wilson chain. 
The parameters $\epsilon_n$ and $t_n$ are computed from $\Delta(\omega)$ using a standard iterative algorithm \cite{bulla_numerical_2008}. 
When $n$ is sufficiently large, there is always $t_n \to \frac12 (1+\Lambda^{-1}) \Lambda^{-\frac12 (n-1)}$ and $\epsilon_n \sim \Lambda^{-n}$. 
Therefore, the site index represents a logarithmic energy scale of the single-impurity problem, and $\hH_{ \infty}$ faithfully describes the low energy physics of $\hH_{\rm SI}$. 
To approach $\hH_{\infty}$, one can define the scaled Hamiltonian's as $\widetilde{H}_N = (\Lambda)^{\frac12 N-1} \hat{H}_N$ and construct them iteratively 
{\small
\begin{align} \label{eq:NRG-main}
\widetilde{H}_{N+1} =& \Lambda^{\frac12} \widetilde{H}_N + \Lambda^{\frac12(N-1)} \sum_{\alpha s}\big( \epsilon_{N+1} d_{N+1,\alpha s}^\dagger d_{N+1,\alpha s} \nonumber\\
& + t_N d_{N+1,\alpha s}^\dagger d_{N,\alpha s} + t_N d_{N,\alpha s}^\dagger d_{N+1,\alpha s}  \big)\ .  
\end{align}
}%
The Hilbert space dimension increases exponentially with $N$. 
The NRG algorithm truncates the Hilbert space by keeping a fixed number (chosen to be $\sim$1600 in this work) of the lowest-lying states at each iterative step.
In order to keep the symmetry in the truncated Hilbert space, in practice we keep all the states up to the lowest gap above the 1600th state. 
Two successive transformations that take $\td{H}_{N}$ to $\td{H}_{N+2}$ can be thought as a renormalization group operation \cite{wilson_nrg1_1980,wilson_nrg2_1980}. 
The system is said to achieve a fixed point when $\td{H}_{N}$ and $\td{H}_{N+2}$ have the same low-lying many-body spectrum. 
It is worth mentioning that freezing the inactive electrons and deriving the effective models (\cref{eq:H01v,eq:HI1v,eq:HSI}) with two-orbital impurities are crucial to apply the NRG approach; otherwise, the impurity would have four orbitals, and the fast increase of the Hilbert space dimension is beyond the scope of NRG. 

In \cref{fig:phase}(c)(d) and \cref{app:flow-diagram} in \cref{app:flow} we plot the lowest many-body levels of the scaled Hamiltonian $\td{H}_N$ ($N\in$ odd) with different parameters \{$\ee_f$, $\ee_{c,1}$, $\ee_{c,3}$\}. 
Due to the U(4) symmetry of $\hH_{\rm SI}$, all the many-body levels can be classified into symmetry sectors labeled by ($Q;\rho$), where $Q$ is the total U(1) charge and $\rho = [m_1,m_2\cdots]$ is the SU(4) representation (Young tableau notation).  
Here we take the convention that $Q=0$ corresponds to a total occupation $2N+2$ ($2N$) for odd (even) $N$.
A fixed point is achieved if the spectrum remains unchanged with $N$, such as the last ten steps in \cref{fig:phase}(c). 
Low energy physics such as spin-susceptibility and spectral density are determined by the many-body levels at the fixed point. 
Readers may refer to Wilson's original papers \cite{wilson_nrg1_1980,wilson_nrg2_1980} for a complete discussion about the fixed points. 

For our model, we find two distinct types of fixed points:
(i) the strong coupling fixed point exhibiting a Fermi liquid behavior, \eg \cref{fig:phase}(c), \cref{app:flow-diagram}(a) and (ii) the LM fixed points exhibiting nonzero U(4) moments, \eg \cref{fig:phase}(d),\cref{app:flow-diagram}(b)(c). 
At a strong coupling fixed point, for either even or odd $N$, the ground state is a singlet with $Q=4k$, $\rho=[0]$ for some order one integer $k$, which in most cases equals 0. 
The low-lying many-body spectrum is identical to the one of a free-fermion chain as if the impurity was nonexistent.
The strong coupling fixed points can be further divided into the frozen impurity regime and the Kondo regime. 
In the frozen impurity regime, the impurity is effectively empty or full and does not enter the low energy physics. 
In the Kondo regime, the impurity forms an LM but is screened by the bath electrons around it, hence the low energy physics is dominated by the free-fermion sites that are effectively decoupled from the screened impurity. 
For example, in \cref{fig:phase}(c), the ground states at the first few steps are not the singlet state, implying that the bath electrons have not yet completely screened the LM. 
After the seventh step, the ground states become the singlet state as the LM is screened, and the spectrum eventually becomes the same as a free-fermion system when the fixed point is achieved. 
To be concrete, we can understand the lowest four-fold ($Q;\rho$)=(-1;[111]) level ($E_h$) and the lowest four-fold (1;[1]) level ($E_e$) as the minimal hole and particle excitations, respectively, and other low energy states as multiple hole and particle excitations. 
For example, the four-fold (-3;[1]) level and the six-fold (-2;[11]) level have the energies $3E_h$ and $2E_h$, respectively, and they can be understood as non-interacting three-hole and two-hole excitations. 
The one-fold (0;[0]) level and the fifteen-fold $(0;[211])$ level have the energy $E_h+E_e$, and they can be understood as non-interacting particle-hole excitations. 
Correspondingly, their SU(4) representations [0]$\oplus$[211] are also given by the direct product $[111]\otimes [1]$ of the representations of the hole and the electron. 
In \cref{app:flow} we show another spectrum in the Kondo regime in \cref{app:flow-diagram}(a). One can verify that its low-lying states at the fixed point are also the same as a free-fermion system.

At an LM fixed point, the low-lying many-body spectrum is identical to a free-fermion chain plus a detached LM.  
They are unstable fixed points if the hybridization function is nonzero at the Fermi level, \ie $\Delta(0)>0$ \cite{wilson_nrg1_1980,wilson_nrg2_1980}. 
However, as shown in \cref{eq:hybridization-main} and \cref{fig:model-main}(d), the single-impurity model $\hH_{\rm SI} $ has $\Delta(0)=0$ if $\ee_{c,3}>0$ and $\ee_{c,1}<0$. 
In this case, the Wilson chain will be broken into two disconnected chains. 
For the parameters used in \cref{fig:phase}(d), $t_5=0$ and the first five bath sites cannot fully screen the impurity. 
Thus, the first six sites, including the impurity, form an effective LM, and the remaining bath sites form a free-fermion chain decoupled from the LM. 
We only show the spectra up to $N=5$ in \cref{fig:phase}(d).
Depending on the representation of the ground state, the LM fixed points can be further classified into LM$_{n}$, where $n=1,2,3$ is the effective impurity occupation.
The LM$_n$ ground states have the charge $Q=4k+n$ and form the same SU(4) representations as the ground states of $\hH_f$ (\cref{eq:Hf}) with $n$ impurity electrons, which are [1], [1,1], and [1,1,1] for $n=$1, 2, and 3,  respectively. 
As the ground states of $\td{H}_5$ in \cref{fig:phase}(d), \cref{app:flow-diagram}(b)(c) form the representations (1;[1]), (2;[1,1]), (3;[1,1,1]), respectively, they are the LM$_{1,2,3}$ states. 

By analyzing the fixed points, we obtain a zero temperature phase diagram of $\hH_{\rm SI}$ (\cref{fig:phase}(e)) in the parameter space of $\epsilon_{c,1}, \eef$ for a fixed $G=\ect-\eco$ ($=12$meV).
As shown in \cref{app:phase}(a), (b), phase diagrams with other values of $G$ are qualitatively same as the one at $G=12$meV. 
For the completeness of discussion, here we let  $\eef$ take values in $[-3.5U_1,0.5U_1]$ and $\eco$ take values in [-60meV,0] such that the mean-field values of $\eef$, $\eco$ (\cref{fig:phase}(a)), which are represented by the black trajectory in \cref{fig:phase}(e), are covered in this phase diagram. 
For $-G<\ee_{c,1}<0$, there is $\Delta(0)=0$ and, according to the last paragraph, the ground state belongs to the LM phase if the impurity is nether empty nor full, \ie $-3U_1<\ee_f<0$, and the frozen impurity phase otherwise. 
Starting from an LM$_n$ phase, lowering $\epsilon_{c,1}$ to a value below $-G$ will drive the system into a strong coupling phase due to the finite hybridization.
Phase boundaries between LM phases and the strong coupling phase are indicated by the white lines. 
The strong coupling phase is further divided into a Kondo regime and a frozen impurity regime. 
Later we will determine the crossover boundary between the two regimes, indicated by the dashed lines in \cref{fig:phase}(e), using the spectral density. 
In the Kondo regime, the color in the phase diagram maps the Kondo energy scale determined from the spectral density.

As discussed in \cref{sec:HSI}, if the Hund's coupling $J_H$ ($\approx 0.3$meV) is considered, the U(4) symmetry of $\hH_{\rm SI}$ will be reduced to U(2)$\times$U(2), where the first (second) U(2) subgroup is the spin-charge rotation symmetry within the first (second) orbital.  
Irreducible representations of U(2)$\times$U(2) are labeled by two U(1) charges $Q_1,Q_2$ and two spin moments $S_1, S_2$. 
Turning on $J_H$, although the four-fold LM$_{n=1,3}$ states are now re-labeled by two spin-$\frac12$ doublets, {\it i.e.}, $(Q_1,Q_2;S_1,S_2)=(2k+n,2k; \frac12, 0)$ and $(2k,2k+n; 0, \frac12)$, they stay degenerate in energy with each other due to crystalline symmetries.
On the contrary, the six-fold LM$_2$ states split into a four-fold multiplet $(2k+1,2k+1;\frac12,\frac12)$, which is free from the Hund's coupling, and a two-fold multiplet $(2k+2,2k;0,0)\oplus (2k,2k+2;0,0)$, whose energy is raised by $J_H$. 
Numerical results with finite $J_H$ are given in \cref{app:phase}(c). 

It is also helpful to look at the representations of the LM$_n$ states under the global U(2) symmetry, which acts on the two orbitals with the same spin-charge rotation.
The total charge and spin of LM$_{1,2,3}$ are 1, 2, 3 (mod 4) and $\frac12$, $\frac12\pm \frac12$, $\frac12$, respectively.

\subsection{Spectral density}
\label{sec:spectral}

We calculate the spectral density of the $f$-electrons, $A_{\alpha s} (\omega,T) = -\frac1{\pi} \mathrm{Im} [G_{\alpha s}(\omega,T)]$, with $G_{\alpha s}(\omega,T)$ being the retarded Green's function of $f_{\alpha s}$ at the temperature $T$.
$G_{\alpha s}(\omega,T)$ is given by 
\begin{equation}
    G_{\alpha s} (\omega,T) = \frac{1}{\omega - \eef - \Sigma^0(\omega) - \Sigma^U_{\alpha s}(\omega,T)}
\end{equation}
where $\Sigma^0 (\omega) = \frac1{\pi} \int  d\ee \frac{\Delta(\epsilon)}{\omega + i 0 ^+ - \ee}$ is the single-particle self-energy contributed by the coupling to the bath, and $\Sigma^U_{\alpha s}$ is the correlation self-energy.
As no symmetry-breaking can happen in the single-impurity model, $A$, $G$, and $\Sigma^U$ should be independent of $\alpha s$.
In the following, we will omit the $\alpha s$ subscript for simplicity.  
As mentioned in \cref{sec:THF}, $f$-orbitals locate in the AA-stacking regions, hence $A(\omega,T)$ corresponds to the STM spectra at the AA-stacking region.
We compute $\Sigma^U$ using the equation of motion \cite{bulla_numerical_1998} within the framework of the reduced density matrix method \cite{hofstetter_generalized_2000}. 
Many-body levels at different RG steps are patched together using the method described in Ref.~\cite{bulla_finite-temperature_2001}.

The fixed point in \cref{fig:phase}(c) is in the Kondo regime, hence, its spectral density exhibits sharp resonance peaks.
The fixed point in \cref{fig:phase}(d) is in the LM phase, and its spectral density is dominated by the upper and lower Hubbard bands. 
One can confirm that the fixed points and spectral densities with other parameters in \cref{app:flow-diagram} in \cref{app:flow} obey the same rule.

We compute the spectral densities for all the data points in the phase diagram in \cref{fig:phase}(e).
For every parameter set we identify a spectral peak (at $\ee$) and measure its half width at half maximum ($\delta$).
In the strong coupling phase, a state is identified as in the Kondo regime if $0\in[\ee-\delta,\ee+\delta]$ and in the frozen impurity regime otherwise. 
The crossover boundaries between the two regimes are indicated by dashed lines in \cref{fig:phase}(e). 
In the Kondo regime, the Kondo temperature can be estimated as $ k_B \tk = \sqrt{\ee^2 + \delta^2}$ \cite{coleman_introduction_2015}. 
$k_B\tk$ is indicated by the color in \cref{fig:phase}(e), and is plotted as a function of $\nu$ (using mean-field parameters) in \cref{fig:phase}(b). 
$k_B\tk$ matches well with the Kondo energy scale $D_K$ estimated by the poor man's scaling. 

\begin{figure*}[t]
\centering
\includegraphics[width=\linewidth]{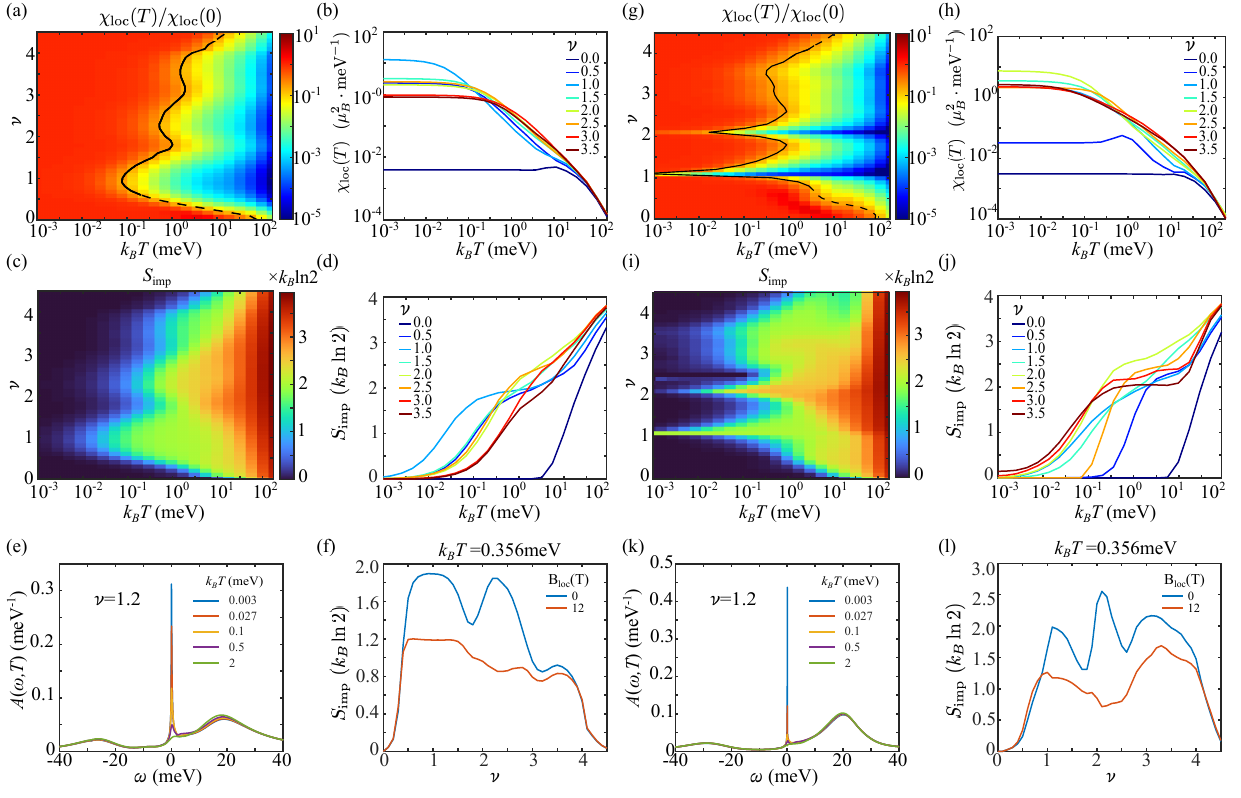}
\caption{Spin susceptibility and entropy contributed by the impurity using single impurity hybridization function (a-f) and DMFT hybridization function (g-l).
(a)(g) $\chi_{\rm loc}(T)/ \chi_{\rm loc}(0)$ as a function of filling $\nu$ and temperature $T$. 
The black curve indicates $\tk$ defined by $\chi_{\rm loc}(\tk)/ \chi_{\rm loc}(0)=1/5 $. Dashed lines are used in the frozen impurity regime.
(b)(h) The local spin susceptibilities $\chi_{\rm loc}(T)$ at fillings $\nu=0, 0.5, 1 \cdots 3.5$. 
(c)(i) The entropy contributed by the impurity as a function of $\nu$ and $T$.
(d)(j) The entropy contributed by the impurity $S_{\rm imp}(T)/(k_B\ln 2)$ at fillings $\nu=0, 0.5, 1 \cdots 3.5$. 
(e)(k) The spectral densities at $\nu=1.2$ at various temperatures. 
(f)(l) The entropy contributed by the impurity as a function of $\nu$ at $B=0,12$ T and temperature $k_BT=0.356 $ meV.
}
\label{fig:chi}
\end{figure*}

\subsection{Local moments and the Pomeranchuk effect}
\label{sec:LM}

At a temperature exceeding the Kondo energy scale, the LM will become effectively decoupled from the bath and visible in experimental measurements. 
This mechanism explains the so-called Pomeranchuk effect \cite{rozen_entropic_2021,saito_isospin_2021} observed in MATBG. 
Refs.~\cite{rozen_entropic_2021} observed a higher entropy ($\sim 1k_B$ per moir\'e cell with $k_B$ being the Boltzmann's constant) state at $\nu\approx 1$ at the temperature $T\approx 10$K.
As this entropy can be quenched by an in-plane magnetic field, it is ascribed to a free local moment. 
Ref.~\cite{saito_isospin_2021} observed a similar effect at $\nu \approx -1$ and showed that an additional resistivity peak that is absent at $T=0$ develops in the higher entropy state at $T\approx 10$K. 
These observations can be naturally explained by the transition from the Fermi liquid phase to the LM phase as the temperature increases.

The LM phase around $\nu\approx 1$ at higher temperatures already manifests itself in the RG flow in \cref{fig:phase}(c). 
At the early stage of the NRG calculation ($N<7$), the ground states form the SU(4) representation $[1]$, which, according to discussions in \cref{sec:NRG}, form the LM$_1$ phase. Only after $N\ge 7$ does the Kondo singlet state [0] cross below the LM states to become the true ground state. Since the NRG iteration can be interpreted as continually lowering the effective temperature, the level crossing during the iteration implies a transition from the strong coupling Kondo phase to the LM phase as temperature increases. 

To further demonstrate the LM phase, we calculate the local spin susceptibilities $\chi_{\rm loc} (T)$ using the filling-dependent $\eco,\eef,G$ parameters given in \cref{fig:phase}(a).
$\chi_{\rm loc} (T)$ is defined as $\frac{d M_{\rm loc}}{d B_{\rm loc}}$ \cite{bulla_numerical_2005,fang_spin_2015} and calculated using linear response theory \cite{fang_spin_2015}, with $M_{\rm loc}$ being the spin moment contributed by the impurity and $B_{\rm loc}$ a local magnetic field that only acts on the impurity. 
As shown in \cref{fig:chi}(a), (b), $\chi_{\rm loc} (T)$ approaches a constant as $T \to 0$, and obeys the Curie's law $\chi_{\rm loc} (T) \sim T^{-1}$, which indicates a free LM at high temperatures. 
One can define the transition temperature ($\tk^s$) between the two behaviors as an alternative estimation of the Kondo temperature. 
Specifically, we find that $\tk^s$ given by $\chi_{\rm loc} (\tk^s) = \frac15 \chi_{\rm loc} (0)$, indicated by the solid black curve in \cref{fig:chi}(a), matches very well with $\tk$ given by the spectral density (\cref{fig:phase}(b)). 
Such determined $\tk^s$ corresponds to the Kondo temperature only in the Kondo regime, where the $f$-orbital is neither empty nor full. In the frozen impurity regime at $\nu$ close to CNP, such determined $\tk^s$ just reflects the energy level of the empty $f$-orbitals and loses the meaning of Kondo temperature. 
Hence we use dashed curve in the frozen impurity regime in \cref{fig:chi}(a). 
Also, one should not confuse this $\tk^s$ with the $\tk$ estimated at CNP (\cref{sec:irrelevance}). The latter is an irrelevant quantity because the RKKY interaction will dominate at CNP and leads to a symmetry-breaking state, based on which the effective models (\cref{eq:H01v,eq:HI1v,eq:HSI}) for $\nu>0$ states are constructed.

We also calculate the impurity entropy $S_{\rm imp}(T)$ for comparison with experiments.
$S_{\rm imp}(T)$ is defined as the difference of the entropy of $\td{H}_N$ and that of a reference free-fermion chain defined by the same $\epsilon_n, t_n$ parameters as in $\td{H}_N$ without the impurity.
As shown in \cref{fig:chi}(c) and (d), $S_{\rm imp}(T)$ is zero in the Fermi liquid phase at sufficiently low $T$ and starts to increase when $T$ reaches the Kondo energy scale. 
For $\nu=1$, $S_{\rm imp}(T)$ climbs to about $\ln 4 \cdot k_B$ - entropy of the four-fold LM$_1$ state - at about $k_B T \approx 0.1$meV and stays around this value until $k_B T$ reaches 2meV. 
The entropy continues to increase for larger $T$ as higher exicted states are involved. 
We also show the temperature-dependent spectral density around $\nu=1$ in \cref{fig:chi}(e).
Consistent with the entropy and spin susceptibility results, the resonance peak dies out for $T>\tk$. 

In \cref{fig:chi}(f), we plot $S_{\rm imp}$ as a function of the filling at a finite temperature under $B_{\rm loc} = 0$ and 12T. 
(See  Fig. 2(e) of Ref.~\cite{rozen_entropic_2021}.)
The entropy with $B_{\rm loc}=0$ has three peaks and two dips. 
Looking at the phase diagram and the $\eco,\eef$ trajectories in \cref{fig:phase}(e), we find that the three peaks correspond to the three domes of LM$_{1,2,3}$, respectively, where the Kondo temperature is relatively lower, and the two dips correspond to the mixed valence states, where the Kondo temperature is higher due to valence fluctuation. 
A finite $B_{\rm loc}$ will polarize the spin and hence suppress the entropy. 
According to the orbital degeneracy, a strong $B_{\rm loc}$ can reduce the entropy at $\nu=1$ to $\ln 2\cdot k_B$.

\section{Solution to the periodic Anderson model}
\label{sec:DMFT}

\subsection{The DMFT+HF approach}
\label{sec:DMFTmodel}
%The NRG calculations consider a single impurity model and neglect the lattice effect. 
 To capture the lattice coherence in the effective periodic Anderson model $\hH^{\rm eff}_0 + \hH^{\rm eff}_I$ (\cref{eq:H01v,eq:HI1v}), we perform a dynamic mean-field \cite{metzner_correlated_1989,georges_dynamical_1996,pruschke_low-energy_2000} decomposition of the on-site interaction $U_1$ and a static HF mean-field decomposition of other interactions. 
This method assumes that spatial correlations are irrelevant in MATBG, which might be justified by the quantum-dot behavior observed in STM.

We assume no symmetry-breaking in the DMFT+HF calculation and will discuss the effect of symmetry-breaking in \cref{sec:DMFT-competition}. As explained in \cref{sec:HSI}, the static mean-fields are then characterized by $\nu_f=\inn{n_{\RR}^f}$ - the occupation of $f$-electrons - and $\nu_{c,a}=\inn{\delta n^c_{\qq=0,a}}$ - the occupations of $c$-electrons. 
There is $\nu_{c,1}=\nu_{c,2}$,  $\nu_{c,3}=\nu_{c,4}$ as crystalline symmetries are assumed. 
Then $\hH^{\rm eff}_0 + \hH^{\rm eff}_I$ can be approximated by 
{\small
\begin{align} \label{eq:H1v_MF}
& \hat{H}^{\rm eff}  \approx \sum_{{\kk s} aa' } \pare{ H_{a a'}^{(c)}(\kk)  +  \delta_{aa'} \ee_{c,a} }c_{\kk a s}^\dagger c_{\kk a' s} + \ee_f \sum_{\RR} n^f_{\RR}
\nonumber \\
& + \sum_{\substack{\kk a\alpha s}} \pare{ e^{-\frac12\lambda^2\kk^2} H^{(cf)}_{a \alpha}(\kk) c^\dagger_{\kk a s}f_{\kk \alpha s} + h.c.} 
+  \frac{U_1}2 \sum_{\RR } :n^f_{\RR}n^f_{\RR}: \ . 
\end{align}}%%
As derived in \cref{sec:HSI}, $\ee_{c,1} = \ee_{c,2} = V_0 \nu_c + W_1 \nu_f - \mu$, $\ee_{c,3} = \ee_{c,4} = V_0 \nu_c + W_3 \nu_f + \frac{J}2 - \frac14 J \nu_{f} - \mu$, $\ee_f = \frac{U_1}2 + 6 U_2 \nu_f + 2 W_1 \nu_{c,1} + 2W_3 \nu_{c,3} - \frac12 J \nu_{c,3} -\mu $, $\nu_c = \sum_{a} \nu_{c,a}$. 
For given $\nu_{c,a}$ and $\nu_f$, \cref{eq:H1v_MF} defines a standard periodic Anderson model that can be addressed using DMFT. 

We first calculate the (retarded) single-particle Green's function by diagonalizing the single-particle part of \cref{eq:H1v_MF}. 
Dynamics on a single $f$-site is described by the local Green's function $G_{\alpha s}^{\rm loc}(\omega)$. It can be formally written as $1/(\omega - \ee_f - \Sigma^0 (\omega) )$ with $\Sigma^0(\omega)$ being the single-particle self-energy.
$\Sigma^0(\omega)$ and the $U_1$ interaction define an Anderson impurity problem with the hybridization function $\Delta(\omega) = - \mathrm{Im} ( \Sigma^0 (\omega) )$, which can be solved by the NRG approach. 
As discussed in \cref{sec:spectral}, the NRG calculation yields a correlation self-energy $\Sigma^U(\omega)$ and an impurity Green's function $G^{\rm imp}(\omega) = 1/(\omega - \ee_f - \Sigma^0 (\omega) - \Sigma^U(\omega))$. 
We then feed $\Sigma^U(\omega)$ into the Dyson's equation in the lattice model and re-calculate the $G^{\rm loc}_{\alpha s}(\omega)$ that defines a new hybridization function.
The DMFT solution to \cref{eq:H1v_MF} for given $\ee_{c,a}$, $\ee_{f}$ is obtained by repeating the above iterative process until $\Sigma^U(\omega)$ converges. The occupations $\nu_f,\nu_{c,a}$ can also be obtained from the lattice Green's function and used to update $\ee_{f},\ee_{c,a}$.
The full self-consistency is reached when $\Sigma^U(\omega)$, $\nu_{c,a}$, and $\nu_{f}$ all converge.

\subsection{Heavy Fermi liquid and Mott insulator}

\label{sec:DMFT-spectral}

\begin{figure}
    \includegraphics[width=\linewidth]{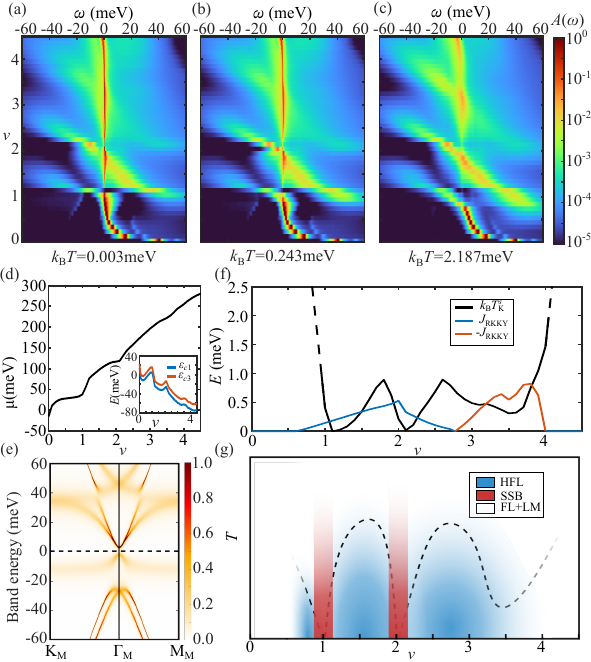}
    \caption{\label{fig:DMFTphase}The DMFT results of spectral densities, chemical potentials, correlated band for Mott insulator, and competition between Kondo effects and RKKY. (a-c) The $f$-spectral densities at filling [0,4.5] at $k_{\rm B}T=0.003,0.243,2.187{\rm meV}$. (d) The chemical potential as a function of filling. Inset: $\epsilon_{c,1},\epsilon_{c,3}$ as a function of fillings. (e) The correlated band for the Mott insulator at $\nu=1.1\approx 1$ at $k_BT=0.003$ meV, which exhibits a gap at the Fermi level and pronounced Hubbard bands. The black dashed line indicates the Fermi level.  (f) Comparison of $\tk$ and $|J_{\rm RKKY}|$. Near $\nu=0$ the state goes into the frozen impurity state, and the half-width of spectral density does not mean Kondo temperature, so a dashed line is used. (g) The finite temperature phase diagram.  The colors indicate the phase, where SSB means spontaneous symmetry-breaking, HFL means heavy Fermi liquid, and FL+LM  means a Fermi liquid with decoupled local moments. The color transitions smoothly since there is no spontaneous symmetry-breaking at finite temperature in two dimensions, and the Kondo-LM transition is not a phase transition but a smooth crossover. The black dashed line indicates the crossover from the heavy Fermi liquid phase to the metallic phase with the local moment, which vanishes gradually when $\nu$ approaches $0$ and increases above $4$, where the ground state goes into the frozen impurity regime.
    }
\end{figure}

The converged $f$-electron spectral densities $A(\omega,\nu)$ in the energy-filling ($\omega$-$\nu$) parameter plane at temperatures $k_BT=0.003,0.243,2.187{\rm meV}$ are shown in \cref{fig:DMFTphase}(a), (b), (c), respectively.  
At the low temperature $k_BT=0.003{\rm meV}$, which can be considered zero, the state at $\nu=0$ is in the frozen impurity regime with an (almost) zero occupation; hence the spectral weight is mainly distributed at positive energy.
As $\nu$ increases, the spectral peak moves to the zero energy and is eventually pinned at the zero energy to form a Kondo resonance. 
This behavior precisely matches STM experiments at low temperatures ($T<1$K)  \cite{nuckolls_strongly_2020,choi_correlation-driven_2021,choi_interaction-driven_2021,oh_evidence_2021}. 
The resonance peak is a robust feature for all fillings except when $\nu$ is close to 1 or 2, indicating that the ground states at generic fillings are the heavy Fermi liquid.   
Using the parameters given in \cref{sec:THF}, the ground state around $\nu=1$ is a Mott insulator with vanishing zero-energy peak (\cref{fig:DMFTphase}(a), (e)), and the ground state around $\nu=2$ is a heavy Fermi liquid with low $\tk$ ($\sim0.1$meV). 
However, using a lightly smaller $U_2$, \eg 0.70meV, the ground states around both $\nu=1,2$ are Mott insulators (see \cref{app:fitU2}(d)(e) in \cref{app:U2}). 
Therefore, we conclude that ground states around $\nu=1,2$ are either Mott insulators or heavy Fermi liquids with low $\tk$, depending on the Hamiltonian parameters. 
The corresponding DMFT band structures at low temperatures for the Mott insulator, which exhibit a gap at the Fermi level, and for heavy Fermi liquids are shown in \cref{fig:DMFTphase}(e) and \cref{fig:DMFTbands}(a),(c), respectively. We also plot $k_B\tk^s$, which is obtained from the susceptibility as explained in \cref{sec:LM}, as a function of $\nu$ in \cref{fig:DMFTphase}(f).
One can see that $\tk^s$ drops to a vanishing value around $\nu=1$ and becomes rather small ($\sim0.1$meV) near $\nu=2$. 
(The small deviation of minimal $\tk^s$ from $\nu=1,2$ is due to numerical errors in evaluating the occupation.)

The evolution of Hubbard bands can also be observed as $\nu$ changes. 
At higher temperatures (\cref{fig:DMFTphase}(b), (c)), the Kondo resonance peaks are smeared by thermal fluctuations, and the evolution of Hubbard bands becomes clearer. 
As $\nu$ increases from 0 to 4, the Hubbard bands periodically pass through the zero energy, matching the cascade of transitions seen in STM experiments at higher temperatures \cite{wong_cascade_2020,choi_interaction-driven_2021}.

We also calculate the chemical potential as a function of the filling, as shown in \cref{fig:DMFTphase}(d).
It leads to the saw-tooth behavior of the inverse compressibility ($\frac{\mathrm{d}\mu}{\mathrm{d}\nu}$) seen in Refs.~\cite{rozen_entropic_2021,wong_cascade_2020,zondiner_cascade_2020,park_flavour_2021,saito_isospin_2021}. Starting from an integer filling, $\frac{{\rm d}\mu}{{\rm d}\nu}$ will first decrease slowly and finally drops nearly to zero before $\nu$ approaches the next integer.
Then, upon $\nu$ crossing the next integer, the inverse compressibility jumps  to a large value. 
This behavior appears periodically as $\nu$ changes from 0 to 4. 
The inset of \cref{fig:DMFTphase}(d) shows the energy of $c$-electrons with respect to the chemical potential, {\it i.e.}, $\ee_{c,1},\ee_{c,3}$. 
It shows a sudden energy jump of the $c$-electrons around every integer filling, which is also referred to as the so-called ``Dirac revival''.  
Both the saw-tooth shape of inverse compressibility and the energy jump of $c$-electrons can be understood through a quantum dot picture. 
As the on-site Coulomb interaction always favors an integer number of $f$ electrons, when doping upon an integer filling, electrons will first occupy $c$-orbitals before $\nu$ reaches the next integer.
The inverse compressibility, proportional to $1/\rho_c$ with $\rho_c$ being the density of states of $c$-electrons, is large at the beginning because $\rho_c$ is small at the band edge of $c$. 
The inverse compressibility decreases with doping because $\rho_c$ increases with doping. 
Upon $\nu$ reaching the next integer, one electron suddenly moves from $c$-orbitals and to $f$-orbitals to save the kinetic energy, leading to the sudden jump of $\ee_{c,a}$ towards the chemical potential.
This jump resets $\rho_c$ to a small value and $\frac{\mathrm{d} \mu}{ \mathrm{d} \nu}$ to a large value.

The resetting of $\rho_c$ to small values around $\nu=1$ (2) is also consistent with the vanishing (small) $\tk$ because it will significantly suppress the hybridization function. 
$\rho_c$ is also reset near $\nu=3$ but to a relatively larger value that still gives rise to a finite $\tk$. 
The resetting depends on the Hartree energies of the $U,W,V$ interactions, hence, it is also possible to obtain Mott insulators at $\nu=2,3$ if another set of parameters is used. 
For example, for fixed $\nu_{c,a}$, $\nu_f$, a smaller $U_2$ will give a smaller $\ee_f$ that is closer to the band edge of $c$-electrons. 
Thus, generally, a smaller $U_2$ leads to a smaller $\rho_c$ and favors Mott insulator phases at integer fillings. 
This is consistent with the observation of the Mott insulator at $\nu=2$ with a suppressed $U_2$ (=0.70meV), as discussed in the first paragraph of this subsection and shown in \cref{app:fitU2}(d)(e) in the \cref{app:U2}. 

Now we compare the DMFT results to the NRG results of the single-impurity model $\hH_{\rm SI}$ discussed in \cref{sec:single-impurity}. 
At a given filling $\nu$, we obtain the hybridization function $\Delta(\omega)$ and $f$-electron energy level $\eef$ from the converged DMFT+HF calculation at $k_B T=0.003$meV. ($\epsilon_{c,a}$ enter the single impurity model implicitly via $\Delta(\omega)$).
This $\Delta(\omega)$ and $\eef$ are different from the bare hybridization (\cref{eq:hybridization-main}) and the HF self-consistent value $\eef$ (\cref{fig:phase}(a)) used in $\hH_{\rm SI}$. 
We then use this $\Delta(\omega)$ and $\eef$ as input for NRG calculations at various finite temperatures, which provide quick estimations for physical observables.   
We show the spin-susceptibility, impurity entropy, and spectral density at various temperatures obtained in such a way in \cref{fig:chi}(g)-(l). 
Comparing \cref{fig:chi}(g)-(l) to (a)-(f), we find that $\hH_{\rm SI}$ already captures features such as Kondo energy scale and entropy. 
The main difference is that $\hH_{\rm SI}$ does not have  Mott insulator states as DMFT results do (\cref{fig:chi}(g)).
Therefore, we conclude that $\hH_{\rm SI}$ is qualitatively correct in the metallic phases. 

We notice that most previous theories attributed the cascade of transition in the STM spectrum at higher temperatures and the saw-tooth feature of the inverse compressibility to spontaneous symmetry breaking. We emphasize that according to the discussion in this subsection, these phenomena already exist without breaking the symmetries. Yet, we will discuss possible symmetry breakings in the next subsection.

\subsection{Competition between Kondo screening and RKKY interaction}
\label{sec:DMFT-competition}
All spatial correlations have been omitted in the DMFT+HF calculation. 
However, local moments at different sites could interact with each other through the Ruderman-Kittel-Kasuya-Yosida (RKKY) interaction and form a coherent order. 
Thus, the competition between the Kondo effect and RKKY interaction must be taken into account to derive a complete phase diagram. 
We have calculated $\tk^s$ using the susceptibility, as shown in \cref{fig:DMFTphase}(f). 
 
Here we use an HF mean-field calculation to estimate the RKKY energy scale, which is described in detail in \cref{app:RKKY}.
We consider a $1\times 2$ super-cell that contains two adjacent LM's, and set their spin alignment as either ferromagnetic or anti-ferromagnetic.
We emphasize that, due to the U(4) symmetry of the THF model, the ``spin order'' here does not necessarily correspond to the physical spin order, but any of its U(4) partners \cite{TBG3}.  
The RKKY energy $J_{\rm RKKY}$ is estimated as the energy gain at each nearest bond of parallel spins. 
We show $|J_{\rm RKKY}|$ in \cref{fig:DMFTphase}(f), where ferromagnetic ($J_{\rm RKKY}>0$) and anti-ferromagnetic ($J_{\rm RKKY}<0$) couplings are represented by blue and yellow curves, respectively.

When $k_B\tk\gg |J_{\rm RKKY}|$, the Kondo effect must win over the RKKY interaction, and hence the ground state is a symmetric heavy Fermi liquid.
On the contrary, when $k_B\tk \ll |J_{\rm RKKY}|$, one may expect a symmetry-breaking ground state.
The Mermin-Wagner theorem states that the correlation length $\xi$ will diverge (remain finite) as $T\to 0$ if $J_{\rm RKKY}$ is ferromagnetic (anti-ferromagnetic). 
Nevertheless, a finite $\xi$ can also open a charge gap if it is much larger than the moir\'e length scale. 
We sketch a phase diagram in the temperature-filling ($T$-$\nu$) parameter space in \cref{fig:DMFTphase}(g).
At $T=0$, apart from the normal Fermi liquid with frozen impurity appearing at $\nu$ close to 0,  the symmetry-breaking ground state wins out around integer fillings $\nu=1, 2$, and likely, if a smaller $U_2$ were to be used, as explained in the last subsection, also around $\nu=3$, while the heavy Fermi liquid forms at all other non-integer fillings.
Upon heating, the symmetry-breaking states change to correlated states with finite spin correlation length $\xi$, but as long as $\xi$ remains much larger than the moir\'e length scale, the charge gap can be preserved. When $T$ is sufficiently high, LM's become fully disordered, and the system enters the metallic LM phase where LM's and the normal Fermi liquid formed by $c$-electrons coexist.
On the other hand, the symmetric heavy Fermi liquid states at non-integer fillings first remain robust upon heating and then continuously evolve to the metallic LM phases when $T$ rises above $\tk$.

Here we only consider the spin (or its U(4) partners) RKKY interaction to demonstrate possible symmetry-breaking states. Other types of symmetry-breaking, such as Chern insulator states at $\nu=1,2,3$ \cite{kang_strong_2019,bultinck_ground_2020,seo_ferromagnetic_2019,vafek_renormalization_2020,TBG4,TBG6} and stripe  states at $\nu=3$ \cite{kang_strong_2019,fang2023_phase}, are also possible. 
It should also be noticed that our periodic Anderson model is based on the assumption of the valley order at CNP, which may become invalid at large fillings. 
Thus, a more complete description should involve valley fluctuations and other types of symmetry-breaking; we leave this for future studies. 

\subsection{Crossover between heavy Fermi liquid and metallic LM states}
\label{sec:DMFT-bands}

\begin{figure*}[ht]
    \centering
    \includegraphics[width=\textwidth]{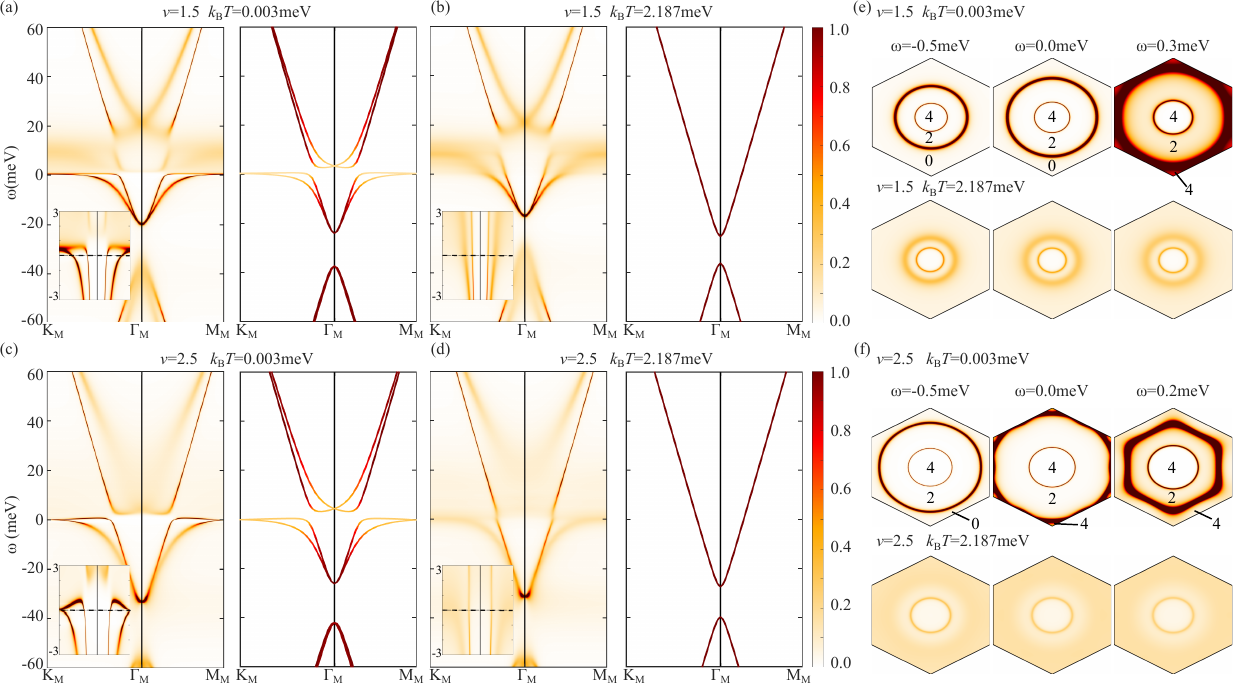}
    \caption{The correlated bands and the energy surfaces. (a-d)The correlated bands of filling $1.5,2.5$ at $k_B{\rm }T=0.003,2.187$meV. Left panel: The $k$ dependent spectral density $A(k,\omega)$ from DMFT along high symmetry path. Inset: zoomed-in diagram for $\omega\in[-3\mathrm{meV},3\mathrm{meV}]$. The zero energy is marked by black dashed lines. Right panel: the bands estimated using a Fermi liquid argument with quasi-particle weight from DMFT results. The color maps the total quasi-particle weight. (e)(f) The $k$ dependent spectral density $A(k,\omega)$ from DMFT in BZ for $\nu=1.5,2.5$ and $k_{\rm B}T=0.003,2.187 {\rm meV}$ at various fixing frequency. The figures at $\omega=0.0$ meV show the Fermi surface. The colormap is the same as (a-d). The numbers between energy surfaces in the low-T plots indicate the number of quasi-particle bands below the associated energy (counted from CNP).
    } 
    \label{fig:DMFTbands}
\end{figure*}

To show the crossover between heavy Fermi liquid and metallic LM states (\cref{fig:DMFTphase}(g)), we plot the total spectral density $A(\kk,\omega)$ at $k_BT=$0.003, 2.187meV and $\nu=1.5$, 2.5 in the left panels of \cref{fig:DMFTbands}(a-d). 
The two temperatures are respectively chosen as much lower and higher than $\tk$, which are 0.994 and 0.857meV for $\nu=1.5$ and 2.5, respectively.  
One can clearly see the heavy $f$-quasi-particle bands at $T\ll \tk$  are smeared out by thermal fluctuations at $T>\tk$.  
The heavy Fermi liquid bands can be understood from a simple Fermi liquid argument. 
At zero temperature, the Landau Fermi liquid theory implies ${\rm Im}\Sigma^U(0)=0$; thus, there are well-defined quasi-particles with long lifetime at low energy.
The main correction is that the $f$-electron is renormalized to $f\approx\sqrt{z}f'$, where $f'$ is the quasi-particle fermion operator and the quasi-particle weight $z$ is given by $z = (1-\partial_{\omega} {\rm Re}\Sigma^U(\omega))^{-1}|_{\omega=0}$.
Then the effective $cf$ hybridization in \cref{eq:H01v} in terms of $f'$ is suppressed by a factor of $z^{\frac12}$.
We obtain $z\approx 0.167,0.271$ at $\nu=1.5,2.5$, respectively, from the self-energy at $k_BT=0.003$meV. 
The estimated heavy Fermi liquid bands obtained from \cref{eq:H01v} with a suppressed $H^{(cf)}$ are shown in the right panels of \cref{fig:DMFTbands}(a), (c).  
The suppression of hybridization results in large effective masses. 
For $T > \tk$, the correlated self-energy $\Sigma^U(\omega)$ has a large imaginary part at zero energy, which means that there is no well-defined $f$-quasi-particle. 
Thus, we have $z=0$, and the $f$-electrons form LM's decoupled from the conduction electrons. The electron band structure is solely contributed by the $c$-electrons, as shown in the right panels of \cref{fig:DMFTbands}(b), (d). 
Comparing to the DMFT bands in the left panel of \cref{fig:DMFTbands}(a-d), the Fermi liquid argument reproduces the heavy quasi-particle bands at $T\ll \tk$ and their disappearance at $T>\tk$.

A characteristic feature of the heavy Fermi liquid is its temperature-dependent Fermi volume \cite{coleman_introduction_2015}. 
Namely, in the Kondo regime, $f$-electrons hybridize with $c$ electrons, hence the area of the Fermi surface counts the total filling. Ideally, at $\nu=1.5$, the total area of two Fermi surface shells, each spin degenerate, should be $0.75S_{\rm BZ}$.
Here $S_{\rm BZ}$ is the area of BZ.
At $\nu=2.5$, one band is fully occupied (\cref{fig:DMFTbands}(c)) and contribute a filling of 2.
The other band has three electron pockets around $\Gamma_M$, $K_M$ and $K_M'$, and their total area should be $0.25S_{\rm BZ}$. 
On the other hand, at $T\gg \tk$ the $f$-electrons do not contribute to electron bands, and the Fermi surfaces should enclose a total area corresponding to the filling $0.5$. 
In the upper and lower panels of \cref{fig:DMFTbands}(e), (f), we plot the DMFT spectral density $A(\kk,\omega)$ at various $\omega$'s around the Fermi level at $T\ll\tk$ and $T> \tk$, respectively. 
The plots at $\omega=0$ sketch the Fermi surfaces, and are consistent with the above analysis. 
 
The evolution of band structures with temperature becomes more evident as we examine the energy dependence of $A(\kk,\omega)$. 
When $T\ll \tk$, due to the heavy band, $A(\kk,\omega)$ exhibits notable change with energy. 
For example, at $\nu=1.5$, because the Fermi level is close to the top of the heavy band (inset of left panel of \cref{fig:DMFTbands}(a)), a hole-type surface will appear when the energy slightly increases, as shown in the upper panel of \cref{fig:DMFTbands}(e).
For another example, at $\nu=2.5$, as $\omega$ decreases, the small electron pockets of the upper band around $K_M$ and $K_M'$ vanish, and then a large hole-type surface in the lower band appear, as shown in the upper panel of \cref{fig:DMFTbands}(f). 
In contrast, the energy surfaces at $T>\tk$ shown in lower panels of \cref{fig:DMFTbands}(e), (f) are almost energy-independent.
Similar energy-dependent behaviors of the energy surfaces occur at other non-integer fillings as well. 
The $T$-dependence of equal-energy surfaces, including the Fermi surface, is a {\it smoking gun} signature of the heavy Fermi liquid phase, and could be observed in spectral experiments such as the quasi-particle interference.

\section{Summary and Discussion}
\label{sec:discussions}

Based on a poor man's scaling analysis, we first argue that the Kondo screening is irrelevant at CNP and the ground state at CNP is a symmetry-breaking state. 
Then, by combining the poor man's scaling, the NRG approach, and the DMFT+HF approach, we have shown that the ground states at $\nu=1,2$ are symmetry-breaking states or Mott insulators and the ground states at most non-integer fillings are heavy Fermi liquids, as summarized in \cref{fig:DMFTphase}(g). 
Upon heating, both of the symmetry-breaking states and the heavy Fermi liquids will eventually evolve into metallic LM states where disordered LM's and a Fermi liquid formed by $c$-electrons coexist. 
In order to verify our theory in future experimental studies, we also predict the temperature-dependence of Fermi volume in the heavy Fermi liquid states at $\nu=1.5$ and 2.5. 

This picture is able to explain several experiments such as STM, transport, the inverse compressibility, \etc, and brings new understandings of the underlying physics.
For example, the spectral density and the Pomeranchuk effect were not connected in previous works. Now our theory states that both of them arise from the Kondo effect. 
The quantum-dot behavior (zero-energy peak) in STM and the higher entropy LM  (lower entropy Fermi liquid) state in compressibility experiment arise from the unscreened LM's (heavy Fermi liquid) at $T>\tk$ ($T<\tk$). 
However, in sharp contrast to the previous understanding that the observed entropy transition  is a first-order phase transition that analogs the liquid-to-solid transition on heating in helium (Pomeranchuk effect), our theory predicts that the transition is instead a {\it continuous crossover}. 

The heavy Fermi liquid states at $\nu=1.5$ and 2.5 are  potential parent states for the observed superconductivity. 
In a naive mean-field picture, the overwhelmingly strong on-site Coulomb repulsion ($U\sim60$meV) would kill any pairing induced by weak attractive interactions, which are merely at the order of 1meV \cite{oh_evidence_2021}. 
However, when the Kondo screening takes place, the $f$-quasi-particles form a heavy Fermi liquid that is free from the strong Coulomb repulsion. 
Instead, the $f$-quasi-particles can only feel a fluctuation-induced residual repulsion $U^*\sim \tk$ \cite{coleman_introduction_2015}, which is at the order of 1meV. 
Then, weak attractive interactions, \eg phonon-mediated \cite{wu_theory_2018,lian_twisted_2019} and spin-wave-mediated interactions, may win over $U^*$ and drive a superconducting phase. 
Therefore, we believe the heavy Fermi liquid phase is a new and useful starting point to study the superconductivity.

{\it Note added.}
During the preparation of the current work, a related work \cite{chou2022kondo} appeared. 
This work studied the symmetric Kondo state using a slave-fermion mean-field in a Kondo lattice model derived from the THF model.  
We are also aware of related works on the Kondo problem in MATBG by A. M. Tsvelik's and B. A. Bernevig's group \cite{BAB2023Kondo,BAB2023Spin}, P. Coleman's group \cite{Coleman2023Kondo}, and a generalization of the THF model to the magic-angle twisted trilayer graphene \cite{yu_2023_magic}. Our results agree with \cite{BAB2023Kondo} that at CNP the ordered state has lower energy than the fully symmetric state. At non-integer fillings, we study the partially symmetric state with the same symmetry as CNP while \cite{chou2022kondo,BAB2023Kondo} consider fully symmetric Kondo states.

\begin{acknowledgements}
We are grateful to B. Andrei Bernevig, Xi Dai, Jia-Bin Yu, Xiao-Bo Lu, Yong-Long Xie, Yi-Lin Wang, and Chang-Ming Yue, Seung-Sup Lee for helpful discussions.  
Z.-D. S., Y.-J W. and G.-D. Z. were supported by
National Natural Science Foundation of China (General Program No.\ 12274005), 
National Key Research and Development Program of China (No.\ 2021YFA1401900).
\end{acknowledgements}

\appendix

\section{More details about the effective Hamiltonian}
\label{app:model}
\subsection{choice of \texorpdfstring{$U_2$}{U2} in the effective single valley model}
\label{app:U2}
\begin{figure}
    \centering
    \includegraphics[width=\linewidth]{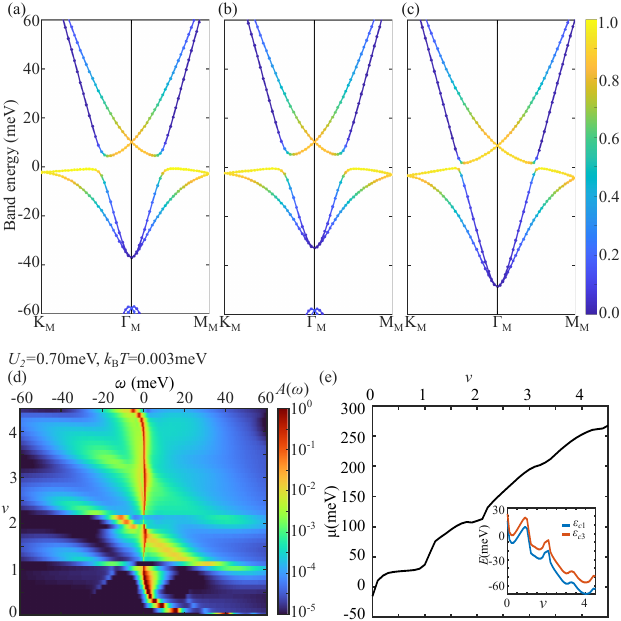}
    \caption{The symmetric HF band at $\nu=4$ and DMFT results with $U_2=0.70\mathrm{meV}$ (a) The  symmetric HF band at $\nu=4$ for the original two-valley model and original parameters. (b) The symmetric HF band at $\nu=4$ for the effective single-valley model with a reduced $U_2=1.16\mathrm{meV}$. (c) The symmetric HF band at $\nu=4$ for the effective single-valley model with original $U_2=2.33\mathrm{meV}$. The color represents the composition of the bands, where yellow and blue correspond to $f$ and $c$ orbitals, respectively. (d) The $f$-spectral density calculated from DMFT at $k_BT=0.003\mathrm{meV}$ with $U_2=0.70\mathrm{meV}$. (e) The chemical potential and $c$-electron mean-field level calculated from DMFT at $k_BT=0.003\mathrm{meV}$ with $U_2=0.70\mathrm{meV}$.}
    \label{app:fitU2}
\end{figure}

We choose a $U_2=1.16\mathrm{meV}$, which is smaller than the original value $U_2=2.33$meV in \cite{song_magic-angle_2022}. This is a compensation for the approximation we made in deriving the effective single valley model. 
As an example, we assume the valley-polarized state as the ground state at CNP. In our effective single valley model upon CNP, we have treated the electrons in the $\eta=+$ valley as a static background. However, once doping away from CNP, the electron will enter or leave $\eta=+$ valley, while the composition of $c,f$-electrons in $\eta=+$ valley will also change.
This gives a correction to the mean field level $\eco,\ect$ and $\eef$ in the single valley effective model for $\eta=-$.
The relative shift between $c$- and $f$-electrons will significantly change the hybridization function at the Fermi level(see \cref{fig:model-main}(d)), while the variation of $G=\epsilon_{c,3}-\epsilon_{c,1}$ plays a minor role (see the comparison of \cref{app:phase}(a)(b)). Therefore,  we neglect the difference between the correction to $\eco,\ect$ for simplicity and only adjust $U_2$, which contributes to Hartree term of $f$-electrons but not $c$-electrons, to mimic the relative shift of $c$- and $f$-electrons energy levels.
We require that at $\nu=4$, where the correction mentioned above reaches the maximum, the conduction band bandwidth of the symmetric self-consistent HF result of the effective single valley is close to the one for $\eta=-$ valley in the two valley model with the original parameters.  
We plot the fully symmetric HF self-consistent bands at $\nu=4$ in \cref{app:fitU2} and find that the band in the effective single valley model with $U_2=1.16$meV(\cref{app:fitU2}(b)) reproduce the one in the original two valley model (\cref{app:fitU2}(a)) better than $U_2=2.33$meV (\cref{app:fitU2}(c)).
From \cref{fig:model-main}(d), one can see that the hybridization increases as $f$ electron rises related to $c$ electron level, which means that reducing $U_2$ leads to a smaller $\tk$ and a stronger quantum-dot-like behavior in general, while the conclusions in \cref{sec:single-impurity,sec:DMFT} are still qualitatively right, as confirmed by \cref{app:fitU2}(d)(e) which shows the spectral density, chemical potential and $c$ electron level for $U_2=0.70\mathrm{meV}$, comparing to those in \cref{fig:DMFTphase} for $U_2=1.16\mathrm{meV}$.

\subsection{Nonzero \texorpdfstring{$M$}{M} term}
\label{app:M}
A generic trial ground state at CNP is given by (\cref{eq:CNP-state})
\begin{equation}
\ket{\Psi_0}= U \prod_{\RR} f_{\RR1+\up}^\dagger f_{\RR1+\down}^\dagger f_{\RR2+\up}^\dagger 
f_{\RR2+\down}^\dagger |{\rm FS}\rangle \ ,
\end{equation}
where $U = \exp(-i\theta_{\mu\nu} \hat{\Sigma}_{\mu\nu}) $ is a U(4) rotation operator and an implicit summation over repeated $\mu,\nu$ indices is assumed.
We can always define the rotated fermion operators 
$\td{c}_{\kk a \eta s} = U c_{\kk a \eta s} U^\dagger$, 
$\td{f}_{\RR a \eta s} = U f_{\RR a \eta s} U^\dagger$ such that $\td{f}_{\RR \alpha + s}$'s are occupied in $\ket{\Psi_0}$ and $\td{f}_{\RR \alpha - s}$'s are empty in $\ket{\Psi_0}$. 
According to the discussions in the supplementary material section S4B of Ref.~\cite{song_magic-angle_2022}, in the flat-band limit ($M=0$), the lowest particle (hole) excitations only involve $\td{c}_{\kk a - s}$ and $\td{f}_{\RR a - s}$ ($\td{c}_{\kk a + s}$ and $\td{f}_{\RR a + s}$). 
Thus, the effective periodic Anderson model for $\nu >0$ derived in \cref{sec:periodic-Anderson} is written in terms of $c_{\kk as} = \td{c}_{\kk a - s}$ and $f_{\RR \alpha s} = \td{f}_{\RR \alpha - s}$. 
Here we give the explicit forms of the rotated operators 
{\small
\begin{equation}\label{eq:feff-def}
\td{f}_{\RR \alpha \eta s} = \sum_{\alpha'\eta' s'} \brak{ e^{i \theta_{\mu\nu} \Sigma^f_{\mu\nu} } }_{\alpha \eta s, \alpha'\eta's'} f_{\RR \alpha' \eta' s'}\ ,
\end{equation}
and
\begin{equation}\label{eq:c12eff-def}
\td{c}_{\kk a \eta s} = \sum_{ \substack{a'=1,2 \\ \eta' s'}} \brak{ e^{i \theta_{\mu\nu} \Sigma^{c12}_{\mu\nu} } }_{a \eta s, a'\eta's'} c_{\kk a' \eta' s'}\quad (a=1,2),
\end{equation}
\begin{equation}\label{eq:c34eff-def}
\td{c}_{\kk a \eta s} = \sum_{ \substack{a'=3,4 \\ \eta' s'}}  \brak{ e^{i \theta_{\mu\nu} \Sigma^{c34}_{\mu\nu} } }_{a \eta s, a'\eta's'} c_{\kk a' \eta' s'}\quad (a=3,4) ,
\end{equation}
}%%
where the eight-by-eight matrices $\Sigma^f_{\mu\nu}$, $\Sigma^{c12}_{\mu\nu}$,  $\Sigma^{c34}_{\mu\nu}$ are defined in \cref{eq:U4f,eq:U4c12,eq:U4c34}.

The $M$-term in the original basis of the THF model (\cref{eq:H02v}) is 
\begin{equation}
M \sum_{aa'=3,4} \sum_{\kk} \sum_{\eta s}
 [\sigma_x]_{aa'} c_{\kk a \eta s}^\dagger c_{\kk a' \eta s}\ .
\end{equation}
It favors the Kramers inter-valley coherent state discussed at the end of \cref{sec:THF}, where $\theta_{x0}$ and $\theta_{y0}$ are nonzero and  satisfy $\theta_{x0}^2 + \theta_{y0}^2=(\pi/4)^2$.
Without loss of generality, we assume $U = \exp(-i\frac{\pi}4 \hat{\Sigma}_{x0})$ for the Kramers inter-valley coherent state. 
Writing this $M$-term in terms of the rotated operators, we obtain
\begin{align}
  M & \sum_{\kk} \sum_{a,a'=3,4 } \sum_{\eta \eta' ss'} \td{c}_{\kk a \eta s}^\dagger O_{a \eta s, a' \eta' s'} \td{c}_{\kk a' \eta' s'} \ ,
\end{align}
where $O = e^{i\frac{\pi}4\Sigma^{c34}_{x0}} \sigma_x \tau_0 \varsigma_0 e^{- i \frac{\pi}4 \Sigma^{c34}_{x0}} = - \sigma_z \tau_x \varsigma_0$. 
The $\tau_x$ matrix in $O$ represents couplings between the empty and occupied single-particle states. 
If we simply project this $M$-term onto the empty states, it vanishes, \ie $[O_{a-s,a'-s'}]=0$. 
A better approximation is applying a Schrieffer-Wolff transformation to decouple the $\eta=\pm$ states, leading to a second-order correction to the effective Hamiltonian. 
As $\inn{\Psi_0| \td{f}_{\alpha \eta s}^\dagger \td{f}_{\alpha\eta s}| \Psi_0} = (1+\eta)/2$, the $J$ term in \cref{eq:HI2v} yields the following mean-field term (see also the supplementary material section S4B of Ref.~\cite{song_magic-angle_2022})
\begin{equation}
    - \frac{J}2 \sum_{a=3,4 } \sum_{\eta  s} \eta \cdot \td{c}_{a\eta s}^\dagger \td{c}_{a\eta s} 
\end{equation}
Then, regarding the $\frac{J}2$ term as the zeroth order Hamiltonian and $M$ as a perturbation, a Schrieffer-Wolff transformation leads to the correction
\begin{equation}
    - \frac{M^2}{J} \sum_{|\kk|<\Lambda_c} \sum_{a=3,4 } \sum_{\eta s} \eta \cdot  \td{c}_{\kk a \eta s}^\dagger \td{c}_{\kk a \eta s} + \mathcal{O}(M^4)\ . 
\end{equation}
The resulting energy levels $ \pm (J/2 + M/J^2) $ at $\kk=0$ is fully consistent with a Taylor expansion of the one-shot energy levels $\pm \sqrt{J^2/4 + M^2}$ derived in Ref.~\cite{song_magic-angle_2022}. 
Projecting the correction to the active d.o.f., \ie $c_{\kk a s} = \td{c}_{\kk a - s}$, we obtain the correction to the effective Hamiltonian
\begin{equation}
    \frac{M^2}{J} \sum_{|\kk|<\Lambda_c} \sum_{a=3,4 } \sum_{s}  {c}_{\kk a s}^\dagger {c}_{\kk a s} + \mathcal{O}(M^4)\ . 
\end{equation}

\subsection{Hund's coupling}
\label{app:Hund}
In this section, we discuss the effect of ferromagnetic coupling $\hH_J$ between $c,f$ electron. We show that besides the Hartree mean field in \cref{eq:Hf,eq:Hc}, it only brings a rather small Hund coupling $J_H$ of $f$ electron; thus, we choose $J_H=0$ and only treat $\hH_J$ by mean-field in the main text. 

Both $\hH^{(cf)}$ and $\hH_J$ couple the $f$-electrons to $c$-electrons, but thanks to the $C_{3z}$ symmetry, the baths they coupled belong to different angular momenta and are independent, for which we can treat them separately. 
In a polar coordinate, $\hH^{cf}$ only couples $f$-electrons to 
\begin{equation}
    \td{c}_{k \alpha s} = \frac1{\mathcal{A}} \sum_{a} \int d\phi \cdot H_{\alpha a}^{(cf)}(\kk) c_{\kk a s} \quad (\alpha=1,2)\ ,
\end{equation}
where $\kk = k(\cos\phi, \sin\phi)$, and $\mathcal{A}$ is a normalization factor. 
Explicitly, there are 
$\td{c}_{k 1 s} \sim \int d\phi \cdot (\gamma c_{\kk 1 s} - v_\star' k e^{i\phi} c_{\kk 2 s})$ 
and 
$\td{c}_{k 2 s} \sim \int d\phi \cdot (\gamma c_{\kk 2 s} - v_\star' k e^{-i\phi} c_{\kk 1 s})$. 
Under the $C_{3z}$ operation (defined in \cref{sec:THF}), $\td{c}_{k 1 s}$ and $\td{c}_{k 2 s}$ have effective angular momenta 1, -1, respectively. 
On the other hand, $\hH_{J}$ only couples $f$-electrons to 
\begin{equation}
\td{c}_{k a s} = \frac1{\mathcal{A}'} \int d\phi \cdot c_{\kk a s}\quad (a=3,4)\ ,
\end{equation}
where $\mathcal{A}'$ is a normalization factor. 
Both $c_{\kk a s}$ ($a=3,4$) have the effective angular momentum 0 under $C_{3z}$. 
Because $\td{c}_{k a s}$ ($a=1,2$) and $\td{c}_{k a s}$ ($a=3,4$) form different representations of $C_{3z}$, they do not couple to each other by $\hH_c$, hence the $\hH_{\rm hyb}$-bath and the $\hH_J$-bath are indeed independent. 

As a ferromagnetic coupling always flows to zero and becomes irrelevant in low energy physics, we can integrate out the $\hat{H}_J$-bath in a single attempt.
\begin{equation}
\hat{H}_H = J_H \sum_{\alpha} n_{\alpha \up} n_{\alpha \down}\ \label{eq:Hund_app},
\end{equation}
where $J_H$, estimated as 0.34meV as shown later, is the additional energy that two electrons will acquire if they occupy the same orbital.
A nonzero $M$ does not change the form of $\hH_H$. 
If $J=0$, there would be no Hund's coupling $J_H$, and $\hH_{\rm SI}$ would have a U(4) symmetry, as expected in a four-flavor Anderson impurity model without multiplet splitting. 

We now derive the effective Hund's coupling $\hH_H$ \cref{eq:Hund_app} in detail. We start from the free $c$-electrons Hamiltonian.
The four-by-four Hamiltonian matrix $H^{(c)}(\kk) + \Delta H^{(c)}$ in \cref{eq:Hc}, \ie $-v_\star (\sigma_x \otimes \sigma_0 k_x + \sigma_y \otimes \sigma_z k_y) + \mathbb{0}_{2\times 2} \oplus G \sigma_0 + \eco \sigma_0\otimes\sigma_0$, can be diagonalized analytically.
As discussed at the end of the last subsection, to $\mathcal{O}(M^2)$, the $M$ term simply shifts the energy of $a=3,4$ electrons by $M^2/J$. 
Thus, all the analysis below applies to the $M\neq 0$ after $G$ is replaced by $G + M^2/J$. 
We find the energy eigenvalues and wave-functions of the $H^{(c)}(\kk) + \Delta H^{(c)}$ as
\begin{equation} \label{eq:Hc-wave1}
\begin{aligned}
\epsilon_1(\kk)  =& \epsilon_+(\kk) = \frac{\eco+\ect}2 + \sqrt{ \frac{G^2}4 + v_\star^2 \kk^2} \\
u_{1}(\kk) =& \begin{pmatrix}
\sin \frac{\theta_\kk}2 e^{-i\phi_\kk} & 0 & -\cos\frac{\theta_\kk}2 & 0
\end{pmatrix}^T
\end{aligned}\ ,
\end{equation}
\begin{equation}
\begin{aligned}
\epsilon_2(\kk) =& \epsilon_+(\kk) = \frac{\eco+\ect}2 + \sqrt{ \frac{G^2}4 + v_\star^2 \kk^2} \\
u_{2}(\kk) =& \begin{pmatrix}
0 & \sin \frac{\theta_\kk}2 e^{i\phi_\kk} & 0 & -\cos\frac{\theta_\kk}2 
\end{pmatrix}^T
\end{aligned}\ ,
\end{equation}
\begin{equation}
\begin{aligned}
\epsilon_3(\kk) =& \epsilon_-(\kk) = \frac{\eco+\ect}2 - \sqrt{ \frac{G^2}4 + v_\star^2 \kk^2} \\
u_{3}(\kk) =& \begin{pmatrix}
\cos \frac{\theta_\kk}2 e^{-i\phi_\kk} & 0 & \sin\frac{\theta_\kk}2 & 0
\end{pmatrix}^T
\end{aligned}\ ,
\end{equation}
\begin{equation} \label{eq:Hc-wave4}
\begin{aligned}
\epsilon_4(\kk) =& \epsilon_-(\kk) =  \frac{\eco+\ect}2 - \sqrt{ \frac{G^2}4 + v_\star^2 \kk^2} \\
u_{4}(\kk) =& \begin{pmatrix}
0 & \cos \frac{\theta_\kk}2 e^{i\phi_\kk} & 0 & \sin\frac{\theta_\kk}2 
\end{pmatrix}^T
\end{aligned}\ . 
\end{equation}
where 
\begin{equation} \label{eq:theta-k}
    \theta_\kk = \arccos \frac{G/2} { \sqrt{G^2/4 + v_\star^2\kk^2} }
\end{equation}
and $\phi_\kk = \arg(k_x + ik_y)$.

Applying a second-order perturbation in terms of $\hH_J$, we obtained the correction to the Hamiltonian 
{\small
\begin{align} \label{eq:DeltaH-tmp1}
& \Delta \hH = - \frac{J^2}{N_M^2} \sum_I \sum_{\substack{\alpha_1\alpha_2\\ s_1 s_1' s_2 s_2'}} \sum_{\substack{\kk_1,\kk_1'\\ \kk_2,\kk_2'} } 
(f_{\alpha_1 s_1}^\dagger f_{\alpha_1 s_1'} - \frac{\nu_f}4\delta_{s_1 s_1'})  \nonumber\\
& \times (f_{\alpha_2 s_2'}^\dagger f_{\alpha_2 s_2} - \frac{\nu_f}4 \delta_{s_2 s_2'}) \cdot  e^{-\frac{\lambda^2}2 (\kk_1^2 + \kk_1^{\prime2} + \kk_2^2 + \kk_2^{\prime2})} \nonumber\\
& \times \frac{ \inn{ \Psi_0 |  c_{\kk_1' \alpha_1+2 s_1'}^\dagger c_{\kk_1 \alpha_1+2 s_1} | \Psi_I } \inn{ \Psi_I |  c_{\kk_2 \alpha_2+2 s_2}^\dagger c_{\kk_2' \alpha_2+2 s_2'} | \Psi_0 }  }{E_I - E_0}\ ,
\end{align}}%%
where $\ket{\Psi_I}$ are excited states with a single particle-hole pair and $E_I$ are the energies of the excited states. 
$\kk_{1,2}$, $\kk_{1,2}'$ are limited up to the cutoff $\Lambda_c$.
The $\hH_J$ here contains a damping factor  $e^{-\frac{\lambda^2}{2}\left(\kk^2+\kk'^2\right)}$ for $c^\dagger_{\kk'} c_\kk$ term since the localized $f$ electron with spread $\lambda$ does not interact with high energy $c$-electron with $|\kk|\gg \lambda^{-1}$ via $\hH_J$. 
Due to the momentum and spin conservation, for the matrix element to be nonzero, there must be $\kk_1 = \kk_2$, $s_1=s_2$, $\kk_1'=\kk_2'$, $s_1'=s_2'$. 
For simplicity, we rewrite $\kk_1$, $\kk_1'$, $s_1$, and $s_1'$ as $\kk$, $\kk'$, $s$, and $s'$, respectively.
$(\kk,s)$ and $(\kk',s')$ label the particle and the hole excitations, respectively. 
Then the matrix element in the third line of \cref{eq:DeltaH-tmp1} can be written as 
{\small
\begin{align}
&  n_F( \epsilon_-(\kk')) (1 - n_F(\epsilon_+(\kk)))  \nonumber \\
&\times \bra{\Psi_0 } \wick{ \c1 c_{\kk' \alpha_1 +2 s'}^\dagger \c2 c_{\kk \alpha_1 +2 s}  \c2 c^\dagger_{\kk \alpha_2 +2 s} \c1 c_{\kk' \alpha_2 +2 s'}  }  \ket{\Psi_0} 
\end{align} }%%
According to the wave functions given in Eqs. (\ref{eq:Hc-wave1})-(\ref{eq:Hc-wave4}), there are 
$\inn{\Psi_0 | c_{\kk' \alpha_1 +2 s'}^\dagger c_{\kk' \alpha_2 +2 s'} |\Psi_0} = \delta_{\alpha_1\alpha_2} \sin^2\frac{\theta_{\kk'}}2 $, 
$\inn{\Psi_0 | c_{\kk \alpha_1 +2 s} c^\dagger_{\kk \alpha_2 +2 s} |\Psi_0} = \delta_{\alpha_1\alpha_2} \cos^2 \frac{\theta_{\kk}}2$. 
The excitation energy $E_I - E_0$ is given by $\epsilon_+(\kk) - \epsilon_-(\kk')$. 
Thus, $\Delta \hH$ is simplified to 
{\small
\begin{align}
& \Delta \hH = - \frac{J^2}{N_M^2} \sum_{\substack{\alpha s s' \\ \kk \kk'}} 
(f_{\alpha s}^\dagger f_{\alpha s'} - \frac{\nu_f}4\delta_{s s'}) 
(f_{\alpha s'}^\dagger f_{\alpha s} - \frac{\nu_f}4 \delta_{s s'}) \nonumber\\
& \times \frac{ n_F( \epsilon_-(\kk')) (1 - n_F(\epsilon_+(\kk) )) \sin^2\frac{\theta_{\kk'}}2 \cos^2\frac{\theta_{\kk}}2 }{\epsilon_+(\kk) - \epsilon_-(\kk')} \nonumber\\
& \times e^{-\lambda^2(\kk^2 + \kk^{\prime2})}
\end{align}}%%
The $s=s'$ contribution is an effective chemical potential shift, estimated as 0.17meV at CNP, of the $f$-electrons.
As it is much smaller than $U_1$, we will omit the $s=s'$ contribution.
The $s\neq s'$ contribution can be written as 
\begin{equation}
    \hH_{H} = J_H \sum_{\alpha} n_{\alpha \up}^f n_{\alpha \down}^f
\end{equation}
with $J_H$ given by 
{\small
\begin{align}
J_H = &  2 J^2 \pare{ \frac{\Omega_0}{2\pi} }^2 \int_0^{\Lambda_c} dk' \cdot k'
    \int_{k_0}^{\Lambda_c} dk \cdot k \cdot e^{-\lambda^2(k^2 + k^{\prime2})} \nonumber\\
 & \times   \frac{ \sin^2\frac{\theta_{k'}}2 \cos^2\frac{\theta_{k}}2 }{\epsilon_+(k) - \epsilon_-(k')} \ ,
\end{align}}%%
where $k_0 $ is determined by $ \epsilon_+(k_0)= 0$. 
Here we have made use of the fact that $\epsilon_{\pm}(\kk)$ and $\theta_{\kk}$ only depends on $|\kk|$ but not $\phi_\kk$. 
At CNP, $\eco=0$ and $\ect = G = J/2 = 8.19$meV, taking the limit $\Lambda_c\to \infty$, we obtain $J_H\approx 0.34$meV. 
Using the self-consistent values of $\epsilon_{c,a}$ shown in \cref{fig:phase}(a), we find $J_H$ at $\nu=1,2,3,4$ are given by 0.30meV, 0.29meV, 0.28meV, 0.26meV, respectively. 
As $J_H$ is small, in this work, we simply set $J_H=0$ for simplicity.  In \cref{app:phase}(a)(c), we present the NRG phase diagrams with $J_H=0$ and $J_H=0.34$meV, showing that small finite $J_H$ do not lead to qualitative difference. The effect of $J_H$ on the Kondo screening will be discussed in \cref{app:RG-nu123}.

\subsection{Hybridization function}
\label{app:hyb}

By definition, the hybridization function $\Delta(\omega)$ is given by 
\begin{align}
    \Delta(\omega)=\frac{\pi}{N}\sum_{\kk}\sum_n|V_{n\alpha}(\kk)|^2\delta(\omega-\epsilon_{n}(\kk))\label{eq:hyb_def}
\end{align}
where $V_{n\alpha}(\kk)=\sum_{a}u^*_{an}(\kk) H^{(cf)}_{a\alpha}(\kk) e^{-\frac{\lambda^2\kk^2}2}$ is the hybridization between $f_{\alpha s}$ and the $n$-th energy band of $c$-electrons. 
$\Delta(\omega)$ does not depend on $\alpha$ because of the $C_{2z}T$ or $C_{2x}$ symmetry that flips the $\alpha$ index.
Substituting $\epsilon_n(\kk)$ and $u_{a n}(\kk)$ in Eqs.~(\ref{eq:Hc-wave1})-(\ref{eq:Hc-wave4}) into the above equation, we obtain $V_{n\alpha}(\kk)$ for $\alpha=1$ as
{\small
\begin{align}
V_{11} (\kk) =& \gamma \sin\frac{\theta_\kk}2 e^{i\phi_\kk}  e^{-\frac{\lambda^2\kk^2}2} \nonumber\\
V_{21} (\kk) =& v_\star' (-k_x + ik_y) \sin\frac{\theta_\kk}2 e^{-i\phi_\kk} e^{-\frac{\lambda^2\kk^2}2} \nonumber\\
V_{31} (\kk) =& \gamma \cos\frac{\theta_\kk}2 e^{i\phi_\kk} e^{-\frac{\lambda^2\kk^2}2} \nonumber\\
V_{41} (\kk) =& v_\star' (-k_x + ik_y) \cos\frac{\theta_\kk}2 e^{-i\phi_\kk} e^{-\frac{\lambda^2\kk^2}2}\ .
\end{align}}%%
Using the energy eigenvalues in Eqs.~(\ref{eq:Hc-wave1})-(\ref{eq:Hc-wave4}) and the $V_{n\alpha}(\kk)$ matrix elements given above, it is direct to obtain 
{\small
\begin{align} \label{eq:hybridization0}
& \Delta(\omega)=\frac{\Omega_0}{2v_\star^2}\abs{\omega-\frac{\eco+\ect}{2}} \left(\gamma^2+v_\star^{\prime 2}k_F^2\right)e^{-k_F^2\lambda^2}\nonumber\\
& \times \left[\theta(\omega-\ect)\sin^2\frac{\theta_{k_F}}{2}+\theta(\eco-\omega)\cos^2\frac{\theta_{k_F}}{2}\right]
\end{align}}
where $k_F$ is given by 
\begin{align}
    k_F=\frac{1}{v_\star }\sqrt{\left[\omega-{(\eco+\ect)}/{2}\right]^2-\left(G/{2}\right)^2}\ . 
\end{align}
Making use of \cref{eq:theta-k}, \cref{eq:hybridization0} can be further simplified to 
{
\begin{align} \label{eq:hybridization}
\Delta(\omega)=& \frac{\Omega_0}{4v_\star^2} \abs{\omega -\ect} \left(\gamma^2+v_\star^{\prime 2}k_F^2\right)e^{-k_F^2\lambda^2}\nonumber\\
& \times \left[\theta(\omega-\ect)  +\theta(\eco-\omega) \right]\ .
\end{align}}

When $\omega \to \ect^+(\eco-)$, only the first (second) term in the second line of \cref{eq:hybridization} contributes to $\Delta(\omega)$, and $k_F\to 0$.
Then we obtain the asymptotic behavior of
{
$\Delta(\omega)$ as $\omega  \to \ect^+(\eco^-)$ 
\begin{equation} \label{eq:hyb-approx}
\Delta(\omega) \sim
\begin{cases}
\frac{\Omega_0}{4 v_\star^2} \gamma^2 \cdot (\omega  - \ect ) + \mathcal{O}( (\omega  - \ect )^2 ),&\omega  \to  \ect^+ \\
\frac{\Omega_0}{4 v_\star^2} G \gamma^2 + \mathcal{O}(\omega - \eco ), & \omega \to \eco^-
\end{cases}. 
\end{equation} 
}
\section{Poor man's scaling of Anderson models with energy-dependent couplings}
\label{app:RG}

\subsection{Generic theory for \texorpdfstring{U($\NN$)}{U(N)}  models}
\label{app:RG-UN}

We consider the Anderson impurity model with $\NN$ symmetric flavors
{\small
\begin{align}
\hH =& \eef \hat{N}_f + \frac{U}2 \hat{N}_f (\hat{N}_f-1) + 
\sum_{\mu=1}^{\NN} \int_{-D}^D d\ee \cdot \ee \cdot d_{\mu}^\dagger(\ee) d_{\mu}(\ee) \nonumber\\
& + \sum_{\mu=1}^{\NN} \int_{-D}^D d\ee \sqrt{ \frac{\Delta(\ee) }{\pi} }
    ( f_{\mu}^\dagger d_{\mu}(\ee) + h.c.) \ ,
\end{align}}%%
where $\mu$ is the flavor index and $N_f = \sum_{\mu=1}^{\NN} f_{\mu}^\dagger f_{\mu}$. 
We assume the ground state of the isolated impurity has $n_f$ $f$-electrons, which can take values in $1,2,\cdots, (N-1)$. 
(We do not consider the empty case ($n_f=0$), the full case ($n_f=N$), nor the mixed valence case where ground states with different $n_f$ are exactly degenerate.)
We further assume the charge gaps to $n_f-1$ and $n_f+1$ electrons are $\Delta E_-$ and $\Delta E_+ = U - \Delta E_-$, respectively. 
We then apply a Schrieffer-Wolff transformation to obtain an effective Coqblin–Schrieffer model for the Hilbert space restricted to $\hat{N}_f = n_f$
{\small
\begin{align}
\hH =& \sum_{\mu=1}^{\NN} \int_{-D}^D d\ee \cdot \ee \cdot d_{\mu}^\dagger(\ee) d_{\mu}(\ee) + \frac{4 g}{\pi U} \sum_{\mu,\mu'=1}^{\NN} \int_{-D}^D d\ee d\ee'  \bigg[ \nonumber\\
&\qquad  \sqrt{ {\Delta(\ee) \Delta(\ee') } } ( f_{\mu}^\dagger f_{\mu'} - x \delta_{\mu\mu'} )
 d_{\mu'}^\dagger (\ee') d_{\mu} (\ee) \bigg] \ .
\end{align}}%%
Terms that only involve $\hat{N}_f$ are omitted because they only contribute to an energy shift.  
The bandwidth $D$ should be smaller than $\min( \Delta E_+, \Delta E_-)$, otherwise the Schrieffer-Wolff transformation is invalid. 
The parameters $g$ and $x$ are given by 
\begin{equation} \label{eq:g0x0}
    g = \frac{U}4 \pare{ \frac1{\Delta E_+} + \frac1{\Delta E_-} },\qquad
    x = \frac{\Delta E_-}{U}\ ,
\end{equation}
respectively. 
If $\eef = -(n_f - \frac12)U$, there will be $\Delta E_+ = \Delta E_- = \frac12 U$ and $g=1$, $x=\frac12$. 

We now truncate the bandwidth to $D-dD = D( 1- dt)$ ($dt\ll 1$) and consider second-order (in $g$) corrections from the virtual particle ($D-dD<\ee <D$) and hole ($-D<\ee <-D+dD$) excitations. 
The particle excitation contributes to the correction 
{\small
\begin{align} \label{eq:RG-tmp0}
& - \frac{(4g)^2}{(\pi U)^2} \frac1{D} \sum_{\mu_1 \mu_2 \mu_1' \mu_2'} \int_{-D+dD}^{D-dD} d\ee_1 d \ee_2 d \int_{D-dD}^D  d \ee_1' d \ee_2' \nonumber \\
& \times \sqrt{\Delta(\ee_1) \Delta(\ee_2) \Delta(\ee_1') \Delta(\ee_2') } 
d_{\mu_1}^\dagger (\ee_1) \inn{ d_{\mu_1'} (\ee_1') d_{\mu_2'}^\dagger (\ee_2') } d_{\mu_2} (\ee_2) \nonumber \\
& \times ( f_{\mu_1'}^\dagger f_{\mu_1} - x \delta_{\mu_1\mu_1'}) 
    \mathbb{P}
 (f_{\mu_2}^\dagger f_{\mu_2'} - x \delta_{\mu_2 \mu_2'}) \ .
\end{align}
}%%
The denominator $D$ in the factor is the excitation energy of a virtual particle.
$\mathbb{P}$ is a projector to the restricted Hilbert space, where $\hat{N}_f = n_f$. 
The expectation $\inn{ d_{\mu_1'} (\ee_1') d_{\mu_2'}^\dagger (\ee_2') }$ evaluated on the ground state is $\delta(\ee_1'-\ee_2') \delta_{\mu_1'\mu_2'}$.
Then we have 
{\small
\begin{align} \label{eq:RG-tmp1}
& - \frac{(4g)^2}{(\pi U)^2} \frac{dD}{D} \Delta(D) \sum_{\mu_1 \mu_2 \mu'} \int_{-D+dD}^{D-dD} d\ee_1 d \ee_2 \sqrt{\Delta(\ee_1) \Delta(\ee_2) }  \nonumber \\
& \times 
d_{\mu_1}^\dagger (\ee_1) d_{\mu_2} (\ee_2) ( f_{\mu'}^\dagger f_{\mu_1} - x \delta_{\mu_1\mu'})
 (f_{\mu_2}^\dagger f_{\mu'} - x \delta_{\mu_2 \mu'})\ ,
\end{align}}%%
where $\mathbb{P}$ is omitted as it commutes with $f_{\mu_2}^\dagger f_{\mu'}$ and $f_{\mu'}^\dagger f_{\mu_1}$. 
After a few steps of algebra, the four-fermion operator $\sum_{\mu'} f_{\mu'}^\dagger f_{\mu_1} f_{\mu_2}^\dagger f_{\mu'}$ can be simplified to
{\small
\begin{equation} \label{eq:four-fermion-p}
f_{\mu_2}^\dagger f_{\mu_1} + \sum_{\mu'} f_{\mu'}^\dagger f_{\mu'} f_{\mu_1} f_{\mu_2}^\dagger 
= f_{\mu_2}^\dagger f_{\mu_1} (1-n_f) + n_f \delta_{\mu_1 \mu_2}\ ,
\end{equation}}%%
where we have made use of the fact that the Hilbert space is restricted to $\hat{N}_f = n_f$. 
Substituting this into \cref{eq:RG-tmp1}, we obtain the corrections to parameters $g$ and $xg$ as 
\begin{equation}
\frac{d g}{dt} \bigg |_p = \frac{4 \Delta(D(t))}{\pi U}  
\pare{ (n_f-1) + 2 x } g^2 \ ,
\end{equation}
\begin{equation}
\frac{d (xg) }{dt} \bigg |_p = \frac{4 \Delta(D(t))}{\pi U}  
\pare{ x^2 + n_f } g^2 \ .
\end{equation}
Here $t$ is the RG parameter and $D(t) = D e^{-t}$ is the reduced bandwidth after successive $t/dt$ RG steps. 

We then calculate the contribution from virtual hole excitation. 
Following the same process as in the last paragraph, we obtain 
{\small
\begin{align} \label{eq:RG-tmp2}
& - \frac{(4g)^2}{(\pi U)^2} \frac{dD}{D} \Delta(-D) \sum_{\mu_1 \mu_2 \mu'} \int_{-D+dD}^{D-dD} d\ee_1 d \ee_2 \sqrt{\Delta(\ee_1) \Delta(\ee_2) }  \nonumber \\
& \times 
d_{\mu_1} (\ee_1) d^\dagger_{\mu_2} (\ee_2) 
 ( f_{\mu_1}^\dagger f_{\mu'} - x \delta_{\mu_1\mu'})
 \mathbb{P} 
 (f_{\mu'}^\dagger f_{\mu_2} - x \delta_{\mu_2 \mu'}) \nonumber\\
= & \frac{(4g)^2}{(\pi U)^2} dt \Delta(-D) \sum_{\mu_1 \mu_2 \mu'} \int_{-D+dD}^{D-dD} d\ee_1 d \ee_2 \sqrt{\Delta(\ee_1) \Delta(\ee_2) }  \nonumber \\
& \times 
 d^\dagger_{\mu_2} (\ee_2) d_{\mu_1} (\ee_1) 
 ( f_{\mu_1}^\dagger f_{\mu'} - x \delta_{\mu_1\mu'})
 (f_{\mu'}^\dagger f_{\mu_2} - x \delta_{\mu_2 \mu'})\ .
\end{align}}%%
In the second equation, we have omitted an energy constant term from the anti-commutator $\{d^\dagger_{\mu_2} (\ee_2), d_{\mu_1} (\ee_1)\}$. 
$\mathbb{P}$ is omitted in the second equation because it commutes with $f_{\mu'}^\dagger f_{\mu_2} $ and $f_{\mu_1}^\dagger f_{\mu'} $. 
The four-fermion operator $\sum_{\mu'} f_{\mu_1}^\dagger f_{\mu'} f_{\mu'}^\dagger f_{\mu_2}$ can be simplified to $ (\NN - n_f+1) f_{\mu_1}^\dagger f_{\mu_2}$ as the Hilbert space is restricted to $\hat{N}_f = n_f$. 
Then the corrections to $g$, $xg$ from \cref{eq:RG-tmp2} can be read out as
\begin{equation}
\frac{d g}{dt} \bigg |_h = \frac{4 \Delta(-D(t))}{\pi U}  
\pare{ \NN - n_f + 1 - 2x } g^2 \ ,
\end{equation}
\begin{equation}
\frac{d (xg) }{dt} \bigg |_h = \frac{4 \Delta(-D(t))}{\pi U}  
\pare{ - x^2  } g^2 \ .
\end{equation}

Adding up the particle and the hole contributions we can obtain the RG equations for $g$ and $(xg)$. 
The Kondo energy scale $\tk$ can be estimated as the reduced bandwidth $D(t)$ where $g$ diverges. 
For a constant $\Delta(\omega)=\Delta_0$, we obtain 
\begin{equation}
    \frac{d g}{dt} = \frac{4 \Delta_0}{\pi U} \NN g^2,\qquad 
    \frac{d(xg)}{dt} = \frac{ 4 \Delta_0} {\pi U} n_f g^2\ 
\end{equation}
and the solution
\begin{equation}
    g(t) = \frac{g(0)}{ 1 - g(0) \frac{4\Delta_0}{\pi U} \NN \cdot t }\ ,
\end{equation}
\begin{equation}
x(t) = x(0) \frac{g(0)}{g(t)} + \frac{n_f}{\NN} \cdot \frac{g(t)-g(0)}{g(t)}\ .
\end{equation}
where $g(0)$ is the initial condition given in \cref{eq:g0x0}.
$g(t)$ diverges at $t_K = \frac{\pi U}{4 \NN g(0)\Delta_0}$, corresponding the Kondo energy scale $D e^{-t_K} = D e^{- \frac{\pi U}{4 \NN g(0) \Delta_0}} $. 
As $g(t)$ diverges as $t \to t_K $, the second term in $x(t)$ dominates and there must be $x \to \frac{n_f}{\NN}$.
In other words, $x$ flows to the occupation fraction. 

\subsection{Application to the symmetric state at CNP}
\label{app:RG-CNP}
We assume a symmetric state of the THF model at CNP and examine its Kondo energy scale.  The Hamiltonian for the bath and the impurity-bath hybridization here are given by $H^{(c,\eta)}$ and $H^{(cf,\eta)}$ in \cref{eq:H02v}, respectively. The energy eigenvalues and wave-functions of $H^{(c,-)}$ are
{\small
\begin{equation} \label{eq:Hc-cnp-wave1}
\begin{aligned}
\epsilon_1(\kk)  =&  \frac{M}2 + \sqrt{ \frac{M^2}4 + v_\star^2 \kk^2} \\
u_{1}(\kk) =& \frac{1}{\sqrt{2}}\begin{pmatrix}
\sin \frac{\theta_\kk}2 e^{-i\phi_\kk} & \sin \frac{\theta_\kk}2 e^{i\phi_\kk} & -\cos\frac{\theta_\kk}2 & -\cos\frac{\theta_\kk}2
\end{pmatrix}^T
\end{aligned}\ ,
\end{equation}
\begin{equation}
\begin{aligned}
\epsilon_2(\kk) =&  -\frac{M}2 + \sqrt{ \frac{M^2}4 + v_\star^2 \kk^2} \\
u_{2}(\kk) =& \frac{1}{\sqrt{2}}\begin{pmatrix}
\cos \frac{\theta_\kk}2 e^{-i\phi_\kk}  & -\cos \frac{\theta_\kk}2 e^{i\phi_\kk} & -\sin\frac{\theta_\kk}2  & \sin\frac{\theta_\kk}2 
\end{pmatrix}^T
\end{aligned}\ ,
\end{equation}
\begin{equation}
\begin{aligned}
\epsilon_3(\kk) =&  \frac{M}2 - \sqrt{ \frac{M^2}4 + v_\star^2 \kk^2} \\
u_{3}(\kk) =& \frac{1}{\sqrt{2}}\begin{pmatrix}
\cos \frac{\theta_\kk}2 e^{-i\phi_\kk} & \cos \frac{\theta_\kk}2 e^{i\phi_\kk} & \sin\frac{\theta_\kk}2 & \sin\frac{\theta_\kk}2
\end{pmatrix}^T
\end{aligned}\ ,
\end{equation}
\begin{equation} \label{eq:Hc-cnp-wave4}
\begin{aligned}
\epsilon_4(\kk) =&  \frac{M}2 - \sqrt{ \frac{M^2}4 + v_\star^2 \kk^2} \\
u_{4}(\kk) =& \frac{1}{\sqrt{2}}\begin{pmatrix}
 \sin \frac{\theta_\kk}2 e^{-i\phi_\kk} & \sin \frac{\theta_\kk}2 e^{i\phi_\kk} & \cos\frac{\theta_\kk}2  & -\cos\frac{\theta_\kk}2 
\end{pmatrix}^T
\end{aligned}\ . 
\end{equation}}%%
Following \cref{app:hyb}, we obtain $V^{(\eta)}_{n\alpha}(\kk)$ for $\alpha=1,\eta=-$ as
{\small
\begin{align}
V_{11} (\kk) =& \frac{1}{\sqrt{2}}\pare{\gamma e^{i\phi_\kk} - v_\star' k e^{-2i\phi_\kk}}\sin\frac{\theta_\kk}2  e^{-\frac{\lambda^2\kk^2}2} \nonumber\\
V_{21} (\kk) =&  \frac{1}{\sqrt{2}}\pare{\gamma e^{i\phi_\kk}  + v_\star' k e^{-2i\phi_\kk}}\cos\frac{\theta_\kk}2 e^{-\frac{\lambda^2\kk^2}2} \nonumber\\
V_{31} (\kk) =& \frac{1}{\sqrt{2}}\pare{\gamma e^{i\phi_\kk}  - v_\star' k e^{-2i\phi_\kk}}\cos\frac{\theta_\kk}2 e^{-\frac{\lambda^2\kk^2}2} \nonumber\\
V_{41} (\kk) =& \frac{1}{\sqrt{2}}\pare{\gamma e^{i\phi_\kk}  + v_\star' k e^{-2i\phi_\kk}}\sin\frac{\theta_\kk}2 e^{-\frac{\lambda^2\kk^2}2} \ .
\end{align}}%%
where 
\begin{equation}
    \theta_k = \arccos \frac{M/2} { \sqrt{M^2/4 + v_\star^2 k^2} }\ . \label{eq:theta-k-M}
\end{equation}
Using \cref{eq:hyb_def} we obtain the hybridization function contributed by the fully symmetric $c$-bands (\cref{fig:model-main}(b)) for $\alpha=1,\eta=-$ as
{\small
\begin{align} \label{eq:hybridization-M}
& \Delta(\omega)= 
\frac{\Omega_0}{4 v_\star^2}\bigg[ \abs{|\omega|-\frac{M}{2}} \theta(|\omega|-M) \left(\gamma^2+v_\star^{\prime 2}k_{F1}^2\right) 
\sin^2 \frac{\theta_{k_{F1}}}2  \nonumber\\
& \times e^{-k_{F1}^2\lambda^2}  + \abs{|\omega| + \frac{M}2} \left(\gamma^2+v_\star^{\prime 2}k_{F2}^2\right) \cos^2 \frac{\theta_{k_{F2}}}2 e^{-k_{F2}^2\lambda^2} \bigg]\ ,
\end{align}}
where
\begin{equation}
    k_{F1}=\frac{1}{v_\star }\sqrt{\left(|\omega|-M/{2}\right)^2-\left(M/{2}\right)^2},\label{eq:kf1-M}
\end{equation}
\begin{equation}
    k_{F2}=\frac{1}{v_\star }\sqrt{\left(|\omega|+M/{2}\right)^2-\left(M/{2}\right)^2},\label{eq:kf2-M}
\end{equation}
Due to the time-reversal symmetry and crystalline symmetries, as explained in \cref{sec:irrelevance}, the hybridization functions for other $\alpha,\eta$ must be the same. 
We should choose the initial cutoff $D = \frac12 U_1$ beyond which the Schrieffer-Wolff transformation is invalid. 
For these states $k_{F1,2} \lesssim  \frac{U_1}{2v_\star}$ and hence $ v_\star^{\prime 2} k_{F1,2}^2 \lesssim  119.4\mrm{meV}^2$, which is significantly smaller than $\gamma^2 \approx 612.6\mrm{meV}^2$. 
The damping factors $e^{-\lambda^2 k_{F{1,2}}^2} \gtrsim 0.74 $ are also large. 
Thus, in the following we approximate $\Delta(\omega)$ ($|\omega|<U_1/2$) as 
{\small
\begin{align} \label{eq:hybridization-M-approx}
\Delta(\omega)\approx & 
\frac{\Omega_0}{4 v_\star^2}\bigg[ \abs{|\omega|-\frac{M}{2}} \theta(|\omega|-M) \gamma^2 \sin^2 \frac{\theta_{k_{F1}}}2 \nonumber\\
&  \qquad  
+ \abs{ |\omega| + \frac{M}2 } \gamma^2 \cos^2 \frac{\theta_{k_{F2}}}2  \bigg]\ .
\end{align}}%%
\cref{eq:hybridization-M-approx} could be further simplified making use of \cref{eq:kf1-M,eq:kf2-M,eq:theta-k-M}, which yields
\begin{equation}
    \Delta(\omega)=\left\{\begin{array}{cc}
         \frac{2\Delta(0)}{M}|\omega|,&|\omega|>M \\
         \Delta(0) \pare{1+ \frac{|\omega|}{M}},&|\omega|<M
    \end{array}\right.\label{eq:hyb_M_approx}
\end{equation}
where $\Delta(0)=\frac{\Omega_0\gamma^2}{8v_\star^2} M \approx 0.0645M$. 

We first consider the flat-band limit ($M=0$), where $\Delta (\omega) = b |\omega|$ and $b=2\Delta(0)/M\approx 0.1290$.
We also assume that there is no multiplet splitting in the symmetric state such that the effective Anderson model should be a U(8) theory with $n_f=4$. 
Naively applying the RG equations derived in the last subsection gives 
\begin{equation}
\frac{d \td{g}}{dt} = - \td{g} + \frac{4 b D }{\pi U_1} \NN \td{g}^2 ,
\end{equation}
where $\NN=8$, $D=U_1/2$, $\td{g} = g e^{-t}$. 
Due to the particle-hole symmetry at CNP, the initial condition (\cref{eq:g0x0}) is $\td{g}(0) = 1$.
Ostensibly, it seems that there would be an unstable fixed point $\td{g}_* = \frac{2\pi}{4\NN b}$, and an initial $\td{g}(0)$ below (above) it will flow to zero (infinity).
Using the actual parameters, we find $\td{g}_* \approx 1.52$ , hence the system would not be in the Kondo phase. 
This result already differs from the standard case with a constant $\Delta(\omega)$, where a positive $g$ always flows to infinity. 
Furthermore, a more careful RG analysis \cite{fritz_phase_2004} shows that the fixed point $\td{g}_*$ does not really exist, and actually, any positive $\td{g}_*$ flows to a non-Kondo phase. 
It is a false result of the weak coupling expansion, which fails for $\Delta(\omega) \sim |\omega|^r$ with $r > \frac12$.
Thus, a $\Delta(\omega) \sim |\omega|$ bath does not have a strong coupling phase. 
This conclusion is also consistent with numerical studies \cite{chen_kondo_1995,gonzalez-buxton_renormalization-group_1998,ingersent_critical_2002}. 

We then consider the case with $M\neq 0$. 
We use the value $M=3.697$meV, which leads to $\Delta(0)\approx 0.239$meV. 
The RG process can be divided into two stages: $D(t) = \frac12 U_1 e^{-t} > M$ and $D(t) < M$.
The boundary between the two stages is $t_1= \ln \frac{U_1}{2M} $. 
The RG equation in the first stage reads 
\begin{equation}
\frac{dg}{dt} = \frac{2\NN b}{\pi} g^2 e^{-t} \;
\Rightarrow \;  g(t) = \frac{1}{ 1 - \frac{2\NN b}{\pi} (1- e^{-t}) }\ .
\end{equation}
Due to the particle-hole symmetry, the initial condition given by \cref{eq:g0x0} is $g(0)=1$. 
We have 
\begin{equation}
    g_1 = g(t_1) = \frac{1}{ 1 - \frac{2\mathcal{N} b}{\pi} ( 1 - \frac{2M}{U_1}) }
\end{equation}
The RG equation in the second stage is given by 
{\small
\begin{align}
\frac{dg}{dt} =& \frac{4\NN \Delta_0}{\pi U_1} g^2 + \frac{4\NN \Delta_0}{\pi U_1} g^2 \cdot e^{-(t-t_1)} \nonumber \\
\Rightarrow \;  g(t) =& \frac{1}{ g_1^{-1} - \frac{4 \NN \Delta_0}{\pi U_1} (t - t_1) -  \frac{4 \NN \Delta_0}{\pi U_1} (1 - e^{-(t-t_1)}) } \ . 
\end{align}}%%
$g(t)$ diverges at $t_K  \approx t_1 +  \frac{\pi U_1}{ 4 g_1 \NN \Delta_0 } - y $ with $y=1$, corresponding to the Kondo energy scale
\begin{equation}
    D_K = M e^y \cdot e^{- \frac{\pi U_1}{ 4 g_1 \NN \Delta_0 } }
    \approx 3.8 \times 10^{-4} \mrm{meV}. \label{eq:Tk_M_poorman}
\end{equation}

\begin{table}[t]
    \centering
    \begin{tabular}{c|c|c|c}\hline\hline
        $M$(meV) & \makecell[c]{$\tk$(meV)\\ exact $\Delta(\omega)$  }& \makecell[c]{$\tk$(meV) \\ approximated $\Delta(\omega)$ }& $D_K$(meV) \\\hline
        6 & 2.45$\times 10^{-9}$ & 1.43$\times 10^{-9}$ & 5.90$\times 10^{-9}$\\\hline
        7 & 2.61$\times 10^{-8}$ & 1.95$\times 10^{-8}$ & 1.16$\times 10^{-7}$\\\hline
        8 & 1.79$\times 10^{-7}$ & 1.31$\times 10^{-7}$ & 1.10$\times 10^{-6}$\\\hline
        9 & 9.06$\times 10^{-7}$ & 7.42$\times 10^{-7}$ & 6.38$\times 10^{-6}$\\\hline
        10& 3.56$\times 10^{-6}$ & 2.97$\times 10^{-6}$ & 2.64$\times 10^{-5}$\\\hline
        11& 1.15$\times 10^{-5}$ & 8.44$\times 10^{-6}$ & 8.53$\times 10^{-5}$\\\hline
        12& 3.16$\times 10^{-5}$ & 2.39$\times 10^{-5}$ & 2.28$\times 10^{-4}$\\\hline
        13& 7.55$\times 10^{-5}$ & 5.78$\times 10^{-5}$ & 5.28$\times 10^{-4}$\\\hline
        14& 1.65$\times 10^{-4}$ & 1.36$\times 10^{-4}$ & 1.09$\times 10^{-3}$\\\hline
        15& 3.14$\times 10^{-4}$ & 2.72$\times 10^{-4}$ & 2.05$\times 10^{-3}$\\\hline
        \hline
    \end{tabular}
    \caption{Energy scales of a fully symmetric Kondo phase in an $\NN=4$ model at various $M$. This model is used to test the validity of the poor man's scaling formula. 
    The second and third columns are $\tk$ obtained from NRG calculations using the exact hybridization function \cref{eq:hybridization-M} and the approximate one \cref{eq:hyb_M_approx}, respectively. $D_K$ is the Kondo energy scale \cref{eq:Tk_M_poorman} given by poor man's scaling with the approximate hybridization function \cref{eq:hyb_M_approx}. 
    As explained in the text, NRG results for $M\lesssim 4$meV have significant numerical errors and hence are not tabulated here. 
    }
    \label{tab:comparison-M}
\end{table}

To confirm the validity of the two-stage RG procedure, we compare it to the Kondo energy scale $\tk$ given by the NRG spectral density for an $\NN=4$ model defined by the same hybridization function $\Delta(\omega)$.
(Notice that the NRG method does not work well in $\NN=8$ cases due to the large Hilbert space dimension on a site.)
In \cref{tab:comparison-M} we tabulate $\tk$ from exact $\Delta(\omega)$ (\cref{eq:hybridization-M}),  $\tk$ from linearized $\Delta(\omega)$ (\cref{eq:hyb_M_approx}), and the poor man's scaling $D_K$ (\cref{eq:Tk_M_poorman}) at various $M$.
We have chosen the initial cutoff $D$ in the NRG as 100meV and kept $\sim$1600 states every step. 
One can see that (i) the linearization of $\Delta(\omega)$ only slightly suppresses $\tk$, and (ii) $D_K$ is on the same order as $\tk$. 

We do not tabulate $\tk$ and $D_K$ for $M<4$meV because when $M$ is small, our NRG calculations fail to respect the U(4) and particle-hole symmetries due to numerical errors. 
To be concrete, for a large odd $N$ that has reached a fixed point with a singlet ground state, the first excited states should be eight-fold degenerate, containing 4 single-electron and 4 single-hole excitation states. 
However, when $M=3$meV, the energies of these eight excitations are all different, varying from $0.05 D_N$ to $1.06D_N$, where $D_N=D\Lambda^{-(N/2-1)}$ is the scaled cutoff in the $N$-th step. 
A larger $M$ will reduce this splitting. 
For example, the energies of these states range from $0.452D_N$ to $0.616D_N$ for $M=4$meV, and  $0.532D_N$ to $0.534D_N$ for $M=5$meV.
The energy splitting becomes negligible for $M\ge 6$meV. 
This numerical error should arise from the $\omega$-dependence of $\Delta(\omega)$.
Our calculations do not have this problem for a small but constant $\Delta(\omega)$.
Also, our calculations do not have this problem for the $\nu>0$ Kondo phase upon the correlated insulator at CNP discussed in the main text.  

\subsection{Application to the effective model for \texorpdfstring{$\nu>0$}{nu>0} states}
\label{app:RG-nu123}
\begin{figure}
    \centering
    \includegraphics[width=\linewidth]{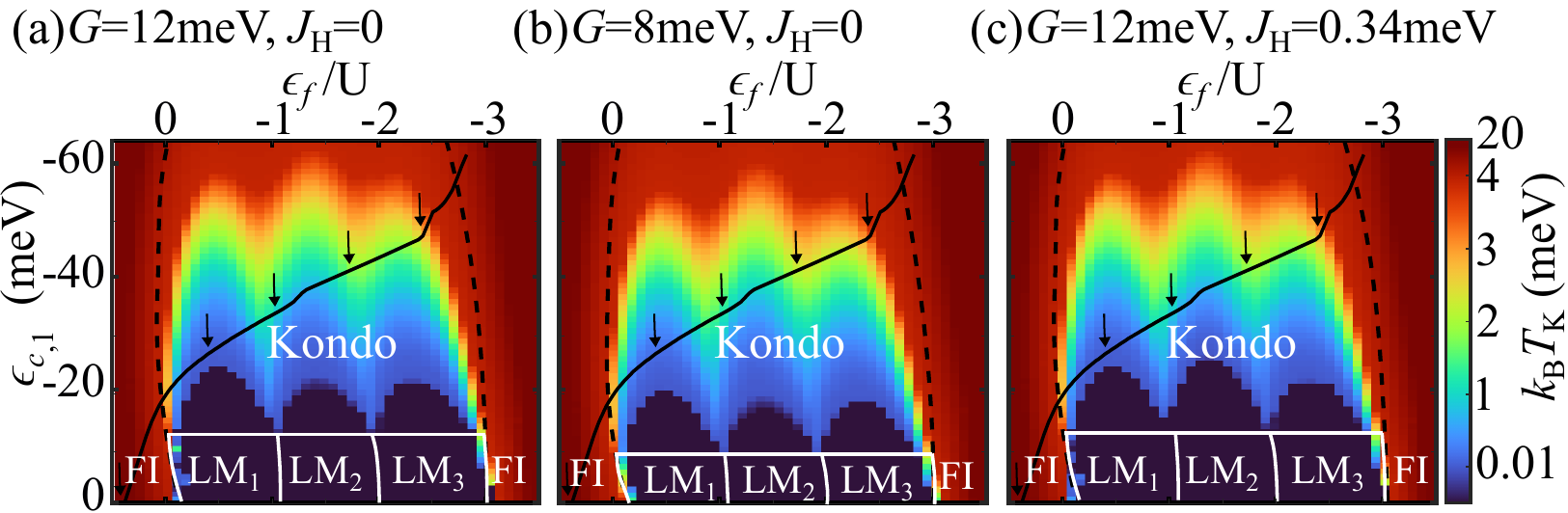}
    \caption{ The phase diagram in the parameter space of $\eco, \eef$ for different $\ect-\eco=G$ and $J_H$. (a) $G=12$meV, $J_H=0$ (same as \cref{fig:phase}(e) in main text), (b) $G=8$meV,$J_H=0$, (c) $G=12$meV,$J_H=0.34$meV. The legend is the same as  \cref{fig:phase}(e). }
    \label{app:phase}
\end{figure}

In the absence of the Hund's coupling $J_H$, we can regard $(\alpha,s)$ as a composite index so that $\hH_{\rm SI}$ (\cref{eq:HSI}) is a U($\NN$) theory with $\NN=4$. 
Then the flow equations in \cref{app:RG-UN} apply. 
For simplicity, we omit the negative branch of $\Delta(\omega)$ (\cref{eq:hybridization}) at $\omega < \eco$ because it is far away from the Fermi level for $\nu>0$ (\cref{fig:model-main}(d)). 
The positive branch of $\Delta(\omega)$ can be well approximated by $\Delta(\omega) = \Delta(0) ( 1 - \omega/\ect )$ for $|\omega| < -\ect $ (\cref{fig:model-main}(d),\cref{eq:hyb-approx}).
We choose the initial cutoff $D$ to be the minimum value of $-\ect$ and $\Delta E_\pm$. 
Substituting this $\Delta(\omega)$ into the general RG equations in \cref{app:RG-UN}, we obtain 
{\small
\begin{equation}\label{eq:flowg}
\frac{dg}{dt} = \frac{4\Delta(0)}{\pi U_1} \NN g^2 + 
\frac{4\Delta(0) D}{-\pi U_1 \ect } ( 4x + 2 n_f - 2 - \NN) g^2 e^{-t}
\end{equation}}%%
and 
{\small
\begin{equation}
\frac{d(xg)}{dt} = \frac{4\Delta(0)}{\pi U_1} n_f g^2 + 
\frac{4\Delta(0)D}{-\pi U_1 \ect } ( 2x^2 + n_f ) g^2 e^{-t}
\end{equation}}%%
The $\mathcal{O}(e^{-t})$ terms will eventually become irrelevant when $t$ is sufficiently large. Since in \cref{eq:flowg} $x$ only enter the $\mathcal{O}(e^{-t})$ terms and is relevant when $t$ is small, we approximate $x$ with its initial value $x=\frac{\Delta E_-}{U}$ (\cref{eq:g0x0}) in \cref{eq:flowg}.
Then the solution of $g$ is 
\begin{equation}
g(t) \approx \frac1{ g^{-1}(0)  - \frac{4 \Delta(0) }{\pi U_1} \NN \pare{ t  + y (1 - e^{-t}) }  }    \ ,
\end{equation}
% $y = \frac{D}{-\ect}  \cdot \frac{3n_f - 6}{\mathcal{N}}$
where $y \approx (\frac{4\Delta E_+}{U_1} + 2 -2n_f) \frac{D}{\NN\ect}$. 
The Kondo energy scale is determined $t = t_K$ at which $g$ diverges.
Assuming $t_K\gg1$, we have 
\begin{equation}
    t_K \approx \frac{\pi U_1}{4 \NN g(0) \Delta(0)} - y
\end{equation}
and hence 
\begin{equation} \label{eq:TK-U4-y}
D_K \approx D \cdot e^{y} \cdot e^{- \frac{\pi U_1}{4 \NN g(0) \Delta(0)}}\ .
\end{equation}

\subsubsection{The \texorpdfstring{$n_f=1,3$}{nf=1,3} cases}

In the presence of the Hund's coupling, we have to examine the derivations in \cref{app:RG-UN} carefully.
The most important effect of $\hH_H$ is to change the local Hilbert space at small energy scales.
In general, $J_H$ leads to a multiplet splitting. 
When the RG energy scale is smaller than the splitting, the higher energy multiplet will become inaccessible, and the local Hilbert space is effectively reduced. 
A minor effect is that the charge gaps $\Delta E_\pm$ will depend on $J_H$ and the resulting coupling between $f$-spin and $d$-spin in the Coqblin–Schrieffer model will break the U($\NN$) symmetry. 

In the following, we study how $\hH_H$ changes the RG equations. 
We first consider the $n_f=1$ case. 
In the virtual particle excitation process (\cref{eq:RG-tmp1}), the intermediate $f$-multiplet is given by $ \ket{F'} = (f_{\mu_2}^\dagger f_{\mu'} - \delta_{\mu_2 \mu'} ) \ket{F}$, where $F$ is the initial $f$-multiplet. 
($\mu$ should be regarded as the composite index $(\alpha,s)$.)
As $\ket{F'}$ has the same particle number as $\ket{F}$, it must be one of the four states with $(n_{1\uparrow},n_{1\downarrow};n_{2\uparrow},n_{2\downarrow})=$ (10;00), (01;00), (00;10), (00;01). 
All of the possible intermediate states do not feel the Hund's coupling ($J_H \sum_{\alpha} n_{\alpha \uparrow} n_{\alpha \downarrow}$) and hence they have the same energy as $\ket{F}$. 
Hence, the excitation energy of the intermediate state is purely contributed by $d$-electrons. 
Then all the following derivations apply. 
The same argument applies to the virtual hole excitation (\cref{eq:RG-tmp2}). 
Therefore, the RG equations for $n_f=1$ will not be affected by $J_H$. 
For the same reason, RG equations for $n_f=3$ will also not be affected by $J_H$, where the initial and intermediate states are single-hole states that do not feel $J_H$. 
The $\tk$ for $n_f=1,3$ is still given by \cref{eq:TK-U4-y}.

\subsubsection{The \texorpdfstring{$n_f=2$}{nf=2} case}

The Hilbert space with two particles has six states: $(n_{1\uparrow},n_{1\downarrow};n_{2\uparrow},n_{2\downarrow})=$ (10;10), (10;01), (01;10), (01;01), (11;00), (00;11).  
The former four states have the energy $2\eef + U_1$, and the latter two states have the energy $2\eef + U_1 + J_H$. 
Thus $J_H$ leads to a multiplet splitting. 
We divide the RG into two stages. 
In the first stage $D(t)$ is larger than $J_H$, then the splitting $J_H$ only plays a minor role and can be neglected.
Thus the RG equations in the first stage are given by \cref{eq:flowg}. 
The first stage ends at $t_1 = \ln (D/J_H)$. 
If $g$ diverges before $t$ reaches $t_1$, the Kondo energy scale should be given by \cref{eq:TK-U4-y} %with $y_2 = 0$, \ie 
\begin{equation}
    D_K ' = D  \cdot e^y \cdot e^{- \frac{\pi U_1}{4 \NN g(0) \Delta(0)}}\ .
\end{equation}
If $g$ is still finite at $t_1$
\begin{equation}
    g_1 =  \frac{g(0)}{ 1  - g(0) \frac{4 \Delta(0) }{\pi U_1} \NN  \left(\ln \frac{D }{J_H}+y\frac{D-J_H}{J_H}\right) } \ ,
\end{equation}
then the RG goes into the second stage. 

The effective cutoff and the initial $g$ of the second stage are $J_H$ and $g_1$, respectively. 
We first examine the virtual particle excitation process (\cref{eq:RG-tmp1}), where the intermediate $f$-multiplet is given by $ \ket{F'} = (f_{\mu_2}^\dagger f_{\mu'} - \delta_{\mu_2 \mu'} ) \ket{F}$.
Here $F$ is the initial $f$-multiplet. 
$\mu',\mu_2$ should be regarded as the composite indices $(\alpha',s')$, $(\alpha_2,s_2)$, respectively.
Suppose $\ket{F}$ is one of the four low energy states, where each orbital ($\alpha=1,2$) has one electron; then, for $\ket{F'}$ to be a low energy state, the index $\mu'$ must have the same orbital index with $\mu_2$, \ie $\alpha'=\alpha_2$, such that each orbital ($\alpha=1,2$) in $\ket{F'}$ still has one electron. 
With this restriction, the four-fermion operator in \cref{eq:four-fermion-p} becomes 
\begin{align} 
f_{\alpha_2 s_2}^\dagger f_{\alpha_1 s_1} + \sum_{ s'} f_{\alpha_2 s'}^\dagger f_{\alpha_2 s'} f_{\alpha_1 s_1} f_{\alpha_2 s_2}^\dagger 
\end{align}
$\sum_{ s'} f_{\alpha_2 s'}^\dagger f_{\alpha_2 s'} $ acting on the bra state (final state) gives $n_{\alpha_2}^f$, which must equal to 1 given that the bra state is one of the four low energy states. 
Thus the four-fermion operator equals to $\delta_{\alpha_2\alpha_1} \delta_{s_2 s_1}$. 
The resulting contributions to the RG equation are 
\begin{equation} \label{eq:RGnf2-tmp1}
\frac{d g}{dt} \bigg |_p = \frac{4 \Delta(D(t))}{\pi U}  
\pare{ 2 x } g^2 \ ,
\end{equation}
\begin{equation} \label{eq:RGnf2-tmp2}
\frac{d (xg) }{dt} \bigg |_p = \frac{4 \Delta(D(t))}{\pi U}  
\pare{ x^2 + 1 } g^2 \ .
\end{equation}
We second examine the virtual hole excitation process (\cref{eq:RG-tmp2}), where the intermediate $f$-multiplet is given by $ \ket{F'} = (f_{\mu'}^\dagger f_{\mu_2} - \delta_{\mu' \mu_2} ) \ket{F}$.
Suppose $\ket{F}$ is one of the four low energy states; then, for $\ket{F'}$ to be in the low energy state, the index $\mu'$ must have the same orbital index with $\mu_2$, \ie $\alpha'=\alpha_2$. 
With this restriction, the four-fermion operator in \cref{eq:RG-tmp2} can be written as 
\begin{equation}
\sum_{s'} f_{\alpha_1 s_1}^\dagger f_{\alpha_2 s'} f_{\alpha_2 s'}^\dagger f_{\alpha_2 s_2}\ .    
\end{equation}
If $\ket{F}$ is one of the four low energy states, it at most occupies one electron in the $\alpha_2$ orbital. 
The $\alpha_2$ orbital of $f_{\alpha_2 s_2} \ket{F}$ must be empty, implying $\sum_{s'} f_{\alpha_2 s'} f_{\alpha_2 s'}^\dagger = 2$. 
Thus the four-fermion operator equals $2 f_{\alpha_1 s_1}^\dagger f_{\alpha_2 s_2} $. 
The resulting contributions to the RG equation are 
\begin{equation} \label{eq:RGnf2-tmp3}
\frac{d g}{dt} \bigg |_h = \frac{4 \Delta(D(t))}{\pi U}  
\pare{ 2 - 2 x } g^2 \ ,
\end{equation}
\begin{equation} \label{eq:RGnf2-tmp4}
\frac{d (xg) }{dt} \bigg |_h = \frac{4 \Delta(D(t))}{\pi U}  
\pare{ - x^2 } g^2 \ .
\end{equation}
\cref{eq:RGnf2-tmp1,eq:RGnf2-tmp2,eq:RGnf2-tmp3,eq:RGnf2-tmp4} are identical to equations of the U(2) case where $\NN=2$, $n_f=1$. 
Following the steps of deriving \cref{eq:TK-U4-y}, we find $x$ still flows to $\frac12$, and 
\begin{equation} 
D_K'' \approx J_H \cdot e^{- \frac{\pi U_1}{8 g_1 \Delta(0)}}\ .
\end{equation}

The final expression for the Kondo energy scale at $n_f=2$ is
\begin{equation}
D_K = \begin{cases}
    D_K^{\prime},\qquad &  D_K' > J_H \\
    D_K^{\prime\prime},\qquad & \text{otherwise}
    \end{cases}\ .
\end{equation}

Several features of  in \cref{app:phase}(a-c) can be understood using the poor man’s scaling result here. First, there are three domes around $\eef=-\frac{1}{2}U_1,-\frac{3}{2}U_1,-\frac{5}{2}U_1$ where $\tk$ is relatively small. They correspond to the $n_f= 1,2,3$ cases here. 
From the poor man’s scaling perspective, these three $\eef$ ’s correspond to the minimal initial value of the coupling constant $g$ (\cref{eq:g0x0}), which leads to smaller $\tk$’s. 
Second, when $J_H=0$ (\cref{app:phase}(a)(b)), for the same $\eco$, the dome at the left has a lower $\tk$ than the dome at the right. This is due to when fixing $\Delta E_+$ a larger $n_f$ means larger $y$ and then larger $\tk$ as argued in \cref{sec:scaling-main}.
Whe $J_H>0$ (\cref{app:phase}(c)) the first dome also has a lower $\tk$ than the third dome since $J_H$ does not affect the $n_f=1,3$ cases.
Third, if $J_H>0$ (\cref{app:phase})(c)), when $|\eco|$ is small ($\lesssim$ 30meV), $\tk$ in the middle dome is the smallest. The reason is that the Kondo energy scale $\tk$ for $n_f = 2$ will be strongly suppressed due to the multiplet splitting if $\tk$ is smaller than $J_H$.
\section{NRG results at other \texorpdfstring{$\epsilon_f,\epsilon_{c1,2},G$}{efec12}}
\begin{figure}
    \centering
    \includegraphics[width=\linewidth]{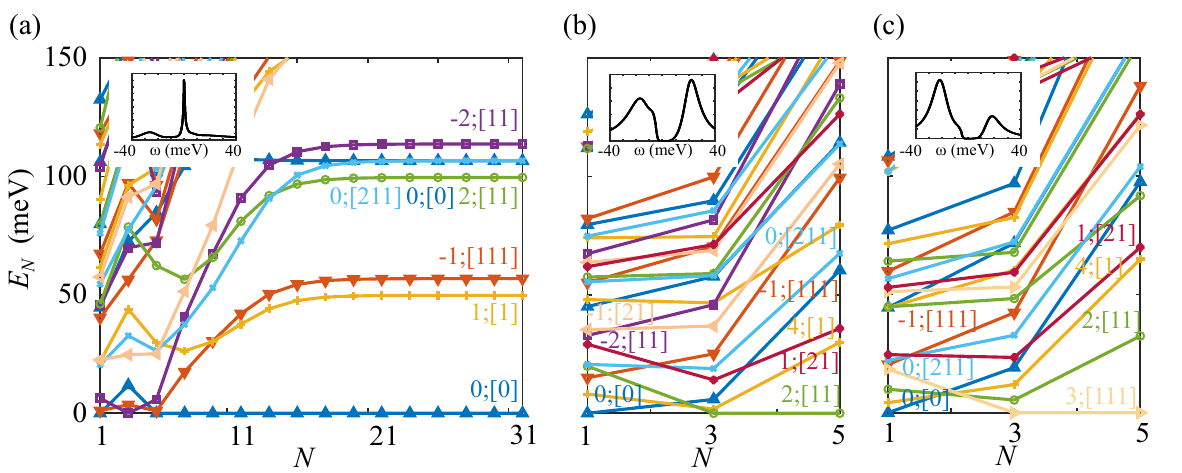}
    \caption{(a) The RG flows of the many-body spectra of the scaled Hamiltonian $\td{H}_N$ ($N\in$ odd) at $\nu=2.5$, with mean-field parameters $\eco=-38.6$meV, $\eef/U_1=-1.39$, $G=12.04$meV.
(b)(c) The RG flows in the LM$_{n}$ phases ($n=2,3$), where $\eef=-(n-\frac12)U_1$, $\eco=-5$meV, and $G=12$meV.
Only spectra at $N<=5$ are shown for the same reason with \cref{fig:phase}(d).
The levels are labeled by total charge $Q$ and the SU(4) irreducible representation. 
The insets are the corresponding spectral densities.
}
    \label{app:flow-diagram}
\end{figure}

Here, we plot the RG flows of the many-body spectra and spectral densities at other $\epsilon_f,\epsilon_{c1,2},G$ in \cref{app:flow-diagram} from the single impurity model as a complement to \cref{fig:phase}(c)(d). The flow diagrams here correspond to the Kondo phase with mean field parameters at $\nu=2.5$ and LM$_{2,3}$ phases, respectively.
\label{app:flow}
\section{Hatree-Fock calculation of RKKY energy scale}
\label{app:RKKY}

In this section, we discuss how we estimate the RKKY energy scale from Hartree-Fock mean field. We consider a $1\times2$ supercell respective to the origin lattice in \cref{fig:model-main}(a), where the order parameters are uniform along $y$ direction but staggered along $\mathbf{a}_1=\frac{2\pi}{3k_\theta}(\sqrt{3},1)$. The supercell is then invariant under translations along  $\mathbf{a}_1'=\frac{4\pi}{3k_\theta}(0,1)$ and $\mathbf{a}_2'=\frac{4\pi}{3k_\theta}(\sqrt{3},0)$, and the state at $\kk$ could couple with the state at $\kk\pm \QQ$ where $\QQ=k_\theta(\frac{\sqrt{3}}{2},0)$. 
The order parameters in real space are defined as 
\begin{align}
    O^{f}_{\alpha s,\alpha's'}(\RR)&=\inn{\Psi|f^\dagger_{\RR\alpha s}f_{\RR\alpha s'}|\Psi}\, ,\nonumber\\
    O^c_{as,a's'}(\RR)&=\frac{1}{N_M}\sum_{\substack{|\kk|,|\kk'|\\<\Lambda_c}} e^{-i(\kk-\kk')\RR}\nonumber\\
    &\left(\langle \Psi | c^\dagger
    _{\kk as}c_{\kk' a's'} | \Psi\rangle-\frac{1}{2}\delta_{\kk\kk'}\delta_{aa'}\delta_{ss'}\right) \; .
\end{align}
where $N_M$ is the number of the original moir\'e unit cell.
The order parameters could be decomposed into a uniform part and a staggered part as $O^f(\RR)=O^{f,0}+e^{i\QQ\cdot\RR}O^{f,1},O^c(\RR)=O^{c,0}+e^{i\QQ\cdot\RR}O^{c,1}$, where

\begin{align}
    O^{f,0}_{\alpha s,\alpha's'}&=\frac{1}{N_M}\sum_{\kk\in \mathrm{MBZ}}\inn{\Psi|f^\dagger_{\kk\alpha s}f_{\kk\alpha's'}|\Psi}\, ,\nonumber\\
    O^{f,1}_{\alpha s,\alpha's'}&=\frac{1}{N_M}\sum_{\kk\in \mathrm{MBZ}}\inn{\Psi|f^\dagger_{\kk+\QQ\alpha s}f_{\kk\alpha's'}|\Psi}\, ,\nonumber\\
    O^{c,0}_{as,a's'}&=\frac{1}{N_M}\sum_{|\kk|<\Lambda_c}\Big(\inn{\Psi|c^\dagger_{\kk as}c_{\kk a's'}|\Psi}-\frac{1}{2}\delta_{\qq,\oo}\delta_{aa'}\delta_{ss'}\Big) \, ,\nonumber\\
    O^{c,1}_{as,a's'}&=\frac{1}{N_M}\sum_{\GG}\sum_{\substack{|\kk|<\Lambda_c,\\|\kk+\QQ+\GG|<\Lambda_c\\}}\inn{\Psi|c^\dagger_{\kk+\QQ+\GG as}c_{\kk a's'}|\Psi}\, ,
\end{align}
where $\GG$ runs over the reciprocal lattice vectors of the unfolded lattice. Whereas, we only keep $\QQ+\GG=\pm\QQ$ for $O^{c,1}$ since other $c^\dagger_{\kk+\QQ+\GG,as}$ create states with large kinetic energy and the corresponding order parameters are small.

For simplicity, we only consider spin density wave and require that the electron density is uniform, i.e. $\nu_f(\RR)=\tr\left[O^f(\RR)\right]$ and $\nu_{c,a}(\RR)=\sum_{s}\left[O^c_{as,as}(\RR)\right]$ do not depend on $\RR$.  The interaction terms in \cref{eq:HI1v} could be decomposed into 
\begin{align}
    \hH^{\rm eff}_I\approx \hH_U^{MF}+\hH_V^{MF}+\hH_W^{MF}+\hH_J^{MF}\nonumber\\
    -E_U-E_V-E_W-E_J 
\end{align}
with their expression listed below as
{\small
\begin{align}
    \hH_U^{MF} &=\sum_{\kk\in \mathrm{fMBZ}}\left[(U_1\nu_f+6U_2\nu_f-U_1O^{f,0}_{\alpha' s',\alpha s} ) \right]\nonumber\\
    &\quad\left(f^\dagger_{\kk \alpha s}f_{\kk \alpha' s'}+f^\dagger_{\kk+\QQ \alpha s}f_{\kk+\QQ \alpha' s'}\right)\nonumber\\
    &\quad-U_1 \left[O^{f,1}_{\alpha' s',\alpha s}f^\dagger_{\kk+\QQ,\alpha s}f_{\kk \alpha' s'}+h.c.\right]\, ,\\
    \hH_V^{MF}&=V_0\nu_c\sum_{as}\sum_{|\kk|<\Lambda_c}c^\dagger_{\kk as}c_{\kk a s}\, ,\\
    \hH_W^{MF}&=\nu_f\sum_{\kk<\Lambda_c}\sum_{as}W_ac^\dagger_{\kk as}c_{\kk as}\nonumber\\
    &\ +\sum_{\kk\in \mathrm{fMBZ}}\sum_{a}W_a\nu_{c,a}\left(f^\dagger_{\kk \alpha s}f_{\kk \alpha' s'}+f^\dagger_{\kk+\QQ \alpha s}f_{\kk+\QQ \alpha' s'}\right),\\
    \hH_J^{MF}&=-J\Bigg\{\sum_{|\kk|<\Lambda_c}\sum_{\alpha ss'}O^{f,0}_{\alpha s',\alpha s} c^\dagger_{\kk,\alpha+2,s}c_{\kk,\alpha+2 s'}\nonumber\\
    &\,+\sum_{\substack{|\kk|,|\kk+\QQ|<\Lambda_c}}\sum_{\alpha ss'}\left(O^{f,1}_{\alpha s',\alpha s} c^\dagger_{\kk+\QQ,\alpha+2,s}c_{\kk,\alpha+2 s'}+h.c.\right)\nonumber\\
    &\hspace*{-1em}+\sum_{\kk\in \mathrm{fMBZ}}\sum_{\alpha ss'}\Big[O^{c,0}_{\alpha+2s,\alpha+2s'}\left(f^\dagger_{\kk\alpha s'}f_{\kk\alpha s}+f^\dagger_{\kk+\QQ\alpha s'}f_{\kk+\QQ\alpha s}\right)\nonumber\\
    & \quad + \left( O^{c,1}_{\alpha+2s,\alpha+2s'} f^\dagger_{\kk+\QQ\alpha s'}f_{\kk\alpha s}+h.c.\right)\Big]\Bigg\}\, ,
\end{align}
}
where fMBZ denotes the folded moir\'e Brioullion and 
{\small
\begin{align}
    E_U&=\frac{U_1}{2}N_M\nu_f^2-\frac{U_1}{2}\sum_{\RR}\tr\left[O^f(\RR)O^f(\RR)\right]+3N_MU_2\nu_f^2\, ,\nonumber\\
    E_V&=\frac{1}{2}V_0N_M\nu_c^2+V_0\sum_{|\kk|<\Lambda_c}4\nu_c\, , \nonumber\\
    E_W&=N_M\sum_a W_a\nu_{c,a}\nu_f+\sum_a\sum_{|\kk|<\Lambda_c}W_a\nu_f\, ,\nonumber\\
    E_J&=-J\sum_{\RR}\sum_{\alpha ss'}O^f_{\alpha s',\alpha s}(\RR)O^c_{\alpha+2 s,\alpha+2 s'}(\RR)\nonumber\\
    &\quad-J\sum_{|\kk|<\Lambda_c}\sum_{\alpha s}\frac{1}{2}O^{f,0}_{\alpha s,\alpha,s}\, .
\end{align}
}
We have neglected the Fock channel of $\hH_V$ as done in S4.A \cite{song_magic-angle_2022}, assuming that symmetry-breaking is mainly contributed by $f$-electron.

We then do two self-consistent calculations to estimate the relative energy between the spin ferromagnetic (FM) state and the anti-ferromagnetic (AFM) state. To be concrete, we choose the spin polarization at the $z$ direction and assume no orbital polarization, which means that $O^f$ is diagonal and $O^f_{1 s,1s}=O^f_{2s,2s}$ for $s=\uparrow/\downarrow$, while $O^c$ is diagonal in spin space and commute with $\sigma_\mu\oplus\sigma_\mu$ for $\mu=x,y,z$ where $\sigma_\mu$ is Pauli matrix act on orbital space. 
We futher enforce $O^{f}(\mathbf{a}_1)=O^{f}(\oo),O^{c}(\mathbf{a}_1)=O^{c}(\oo)$ for FM state and $O^{f}(\mathbf{a}_1)=\varsigma_xO^{f}(\oo)\varsigma_x,O^{c}(\mathbf{a}_1)=\varsigma_xO^{c}(\oo)\varsigma_x$ for AFM state, where $\varsigma_x$ is Pauli $x$ matrix in the spin space. 
We then do the self-consistent calculation as described above, yielding two converged total energy (per moi\'re unit cell) $E_{\rm FM}$ and $E_{\rm AFM}$ for spin FM and AFM state, respectively. We define the RKKY energy scale $J_{\rm RKKY}$ as the effective energy gain at each nearest neighbor bond of parallel spin, where $J_{\rm RKKY}>0 $ means ferromagnetic coupling and $J_{\rm RKKY}<0 $ means anti-ferromagnetic coupling. Then the FM and AFM state gain energy of  $-3J_{\rm RKKY},J_{\rm RKKY}$ per moi\'re unit cell, respectively, for which we can extract  $J_{\rm RKKY}=\frac{1}{4}(E_{\rm AFM}-E_{\rm FM})$. We plot $J_{\rm RKKY}$ as a function of $\nu$ in \cref{fig:DMFTphase}(f).

\bibliography{refs.bib}

\end{document}